\documentclass[showpacs,aps,prd,reprint,superscriptaddress,nofootinbib,longbibliography,preprintnumbers]{revtex4-1}
\usepackage[colorlinks=true, pdfstartview=FitV, linkcolor=magenta,citecolor=blue, urlcolor=magenta,
bookmarks=true, bookmarksnumbered=true, breaklinks]{hyperref}
\usepackage[dvipdfmx]{graphicx}

\usepackage{here}
\usepackage{url}
\usepackage{amsmath,amssymb,bm,color,longtable,mathrsfs,amsfonts,slashed}

\newcommand{\Slash}[1]{{\ooalign{\hfil/\hfil\crcr$#1$}}}

\begin{document}
\begin{flushright}
\end{flushright}

\title{Axial anomaly effect to the chiral-partner structure of diquarks at high temperature}

\author{Daiki~Suenaga}
\email[]{daiki.suenaga@riken.jp}
\affiliation{Few-body Systems in Physics Laboratory, RIKEN Nishina Center, Wako 351-0198, Japan}
\affiliation{Research Center for Nuclear Physics,
Osaka University, Ibaraki 567-0048, Japan }

\author{Makoto~Oka}
\email[]{makoto.oka@riken.jp}
\affiliation{Few-body Systems in Physics Laboratory, RIKEN Nishina Center, Wako 351-0198, Japan}
\affiliation{Advanced Science Research Center, Japan Atomic Energy Agency (JAEA), Tokai 319-1195, Japan}

\date{\today}

\begin{abstract}
Masses of positive-parity and negative-parity diquarks are investigated at finite temperature with a quark chemical potential.
We employ the three-flavor Nambu-Jona-Lasinio model, in order to delineate chiral properties of the diquarks, in particular,
the mass degeneracy of chiral partners under extreme conditions. We focus on the effects of $U(1)_A$ axial anomaly on manifestation of the chiral-partner structures. We find that, in the absence of anomaly effects to the diquarks, the mass degeneracies in all $[ud]$, $[su]$ and $[sd]$ diquark sectors take place prominently above the pseudocritical temperature of the chiral restoration. On the other hand, the anomaly effects are found to hinder the $[ud]$ diquark from exhibiting the mass degeneracy, accompanied by a slow reduction of the $\bar{s}s$ condensate, while the $[su]$ and $[sd]$ diquarks 
are not much affected.
Our present investigation will provide useful information on the chiral-partner structure with the anomaly effects of diquarks for heavy-ion collision experiments of singly heavy baryons and doubly heavy tetraquarks, and for future lattice simulations of the diquarks.
 \end{abstract}


\maketitle

\section{Introduction}
\label{sec:Introduction}

The light diquark, a cluster made of two light quarks ($u$, $d$ and $s$ quarks), is not a direct observable due to the color confinement but is known as a useful ingredient of hadrons. For instance, the mass spectrum and decay properties of singly heavy baryons (SHBs) which consist of one heavy quark ($c$ or $b$ quark) and two light quarks (i.e., one diquark) can be understood from the diquarks dynamics of order $\Lambda_{\rm QCD}$, the typical energy scale of quantum chromodynamics (QCD), since the heavy quark whose mass is larger than $\Lambda_{\rm QCD}$ is regarded as a spectator~\cite{Neubert:1993mb,manohar2000heavy}. Similarly, the diquark is also expected to play an important role in dynamics of delineating doubly heavy tetraquarks such as $T_{cc}$.

Light diquarks contain only light flavors, so that it is useful to classify them from an appropriate chiral-symmetry representation. Focusing on this fact, effective models of diquarks from the viewpoint of chiral symmetry have been constructed~\cite{Hong:2004ux}, and accordingly, investigation of SHBs based on the chiral models have been promoted from field-theoretical approaches with both linear and nonlinear representations~\cite{Ebert:1995fp,Kawakami:2018olq,Kawakami:2019hpp,Harada:2019udr,Dmitrasinovic:2020wye,Kawakami:2020sxd,Suenaga:2021qri,Suenaga:2022ajn} and from diquark-heavy-quark potential description~\cite{Kim:2020imk,Kim:2021ywp,Kim:2022pyq}.

Employing the linear representation of chiral symmetry, we can describe not only ground states but also orbitally excited states carrying opposite parities, known as {\it chiral partners}, in a consistent manner. For diquarks as well, the existence of chiral partners are expected, and theoretically, for instance, properties of $\Lambda_c(1/2^-)$ as a chiral partner of the ground-state 
$\Lambda_c(2286)$ is being explored~\cite{Kawakami:2020sxd}. However, no candidate of such 
$\Lambda_c(1/2^-)$ has been observed~\cite{Workman:2022ynf} and the quest for finding such an excited state is left as an important task.

The mass splitting between the chiral partners is supposed to be generated by the spontaneous breakdown of chiral symmetry. Thus, at extreme conditions such as high temperature and/or density where chiral symmetry may be restored, the masses of the chiral partners will become degenerate~\cite{Hatsuda:1994pi}. 
As a precursor phenomenon, it is expected that the masses of the chiral partners may move towards the degeneracy
at finite temperature and/or density.
If such reduction of the mass splitting of $\Lambda_c(1/2^-)$ and $\Lambda_c(2286)$ is realized in hot and/or dense matter which can be produced in, for example, heavy-ion collision (HIC) experiments, we expect to 
observe a suppression of the $\Lambda_c(1/2^-)$ decay.

Thus far, chiral-partner structures of light mesons~\cite{Hatsuda:1994pi,Chiku:1997va,Harada:2003jx,Grahl:2011yk,GomezNicola:2013pgq,Gubler:2016djf,Suenaga:2019urn,Xu:2021lxa}, nucleons~\cite{Detar:1988kn,Jido:1998av,Zschiesche:2006zj,Motohiro:2015taa,Suenaga:2017wbb} and heavy-light mesons~\cite{Suenaga:2014sga,Sasaki:2014asa,Harada:2016uca,Suenaga:2017deu} have been theoretically investigated. In addition, The partner structures of diquarks and mesons in cold and dense two-color QCD, where the diquarks become color-singlet baryon, have been examined~\cite{Suenaga:2022uqn}, motivated by the applicability of lattice simulations at density~\cite{Hands:2007uc,Murakami:2022lmq}.


In addition to chiral symmetry, QCD possesses another prominent symmetric properties: the $U(1)_A$ axial anomaly~\cite{Weinberg:1975ui}. That is, a conservation law of the $U(1)_A$ axial charge is violated by quantum effects induced by external gluons, which may be understood by instanton effects~\cite{tHooft:1986ooh}. Chiral effective models demonstrate that, at zero temperature and density, the anomaly triggers the {\it inverse mass hierarchy} of negative-parity diquarks~\cite{Harada:2019udr}; a mass of non-strange diquarks $[ud]_-$ becomes larger than that of strange diquarks $[su]_-$ or $[sd]_-$. Such a peculiar inversion indicates that the $U(1)_A$ anomaly effects have a significant influence on the chiral-partner structure of the diquarks in medium.

In light of the above symmetric properties of QCD, in the present study we examine diquark mass changes at finite temperature with a quark chemical potential to see the chiral-partner structures and roles of the $U(1)_A$ axial anomaly for them. In particular, we employ the three-flavor Nambu-Jona-Lasinio (NJL) model incorporating six-point interactions responsible for the $U(1)_A$ axial anomaly effects~\cite{Kobayashi:1970ji,Kobayashi:1971qz,tHooft:1976snw,tHooft:1976rip,Abuki:2010jq}. In hot QCD matter, our present investigation is expected to provide useful information on SHBs and doubly heavy tetraquarks from the viewpoint of chiral symmetry and the $U(1)_A$ axial anomaly for future HIC experiments and lattice simulations. Meanwhile, in cold and dense regime, further understandings of the onset of color superconducting phase~\cite{Buballa:2003qv,Alford:2007xm} and roles of the $U(1)_A$ axial anomaly there, which is related to the continuous transition from hadron to quark phases~\cite{Schafer:1998ef,Hatsuda:2006ps}, are expected.

This article is organized as follows. In Sec.~\ref{sec:Model}, our three-flavor NJL model containing meson and diquark channels is introduced, and in Sec.~\ref{sec:Analysis} our strategy to evaluate the diquark masses in medium and our regularization technique are explained. Based on them, we present numerical results of the diquark masses at finite temperature and chemical potential in Sec.~\ref{sec:Results}. In Sec.~\ref{sec:Discussions}, we discuss the SHB spectrum at finite temperature expected from our results on the diquark masses, and artifacts induced by our regularization. Finally, Sec.~\ref{sec:Conclusions} is devoted to concluding our present study.

\section{Model}
\label{sec:Model}

In this section, we present our NJL model toward investigation of diquark masses at finite temperature with a quark chemical potential.

Our NJL Lagrangian is separated into three parts of
\begin{eqnarray} 
{\cal L}_{\rm NJL} = {\cal L}_{2q} + {\cal L}_{4q} + {\cal L}_{6q}^{\rm anom.}\ . \label{LNJL}
\end{eqnarray}
The first part ${\cal L}_{2q}$ includes kinetic and mass terms of dynamical quarks as
\begin{eqnarray}
{\cal L}_{2q} = \bar{\psi}(i\Slash{\partial}+\mu\gamma_0-{\cal M})\psi\ ,
\end{eqnarray}
where $\psi=(u,d,s)^T$ is a three-flavor quark multiplet. The quantities $\mu$ and ${\cal M}$ are a quark chemical potential and a mass matrix of the current quarks, respectively. Under $SU(2)_I$ isospin symmetry ${\cal M}$ takes the form of ${\cal M} = {\rm diag}(m_q,m_q,m_s)$.

 The second part in Eq.~(\ref{LNJL}), ${\cal L}_{4q}$, describes four-point interactions among the quarks 
\begin{eqnarray}
&& {\cal L}_{4q} = G\sum_{A=0}^8\big[(\bar{\psi}\lambda_f^A\psi)^2 + (\bar{\psi}i\gamma_5\lambda_f^A\psi)^2\big] \nonumber\\
&& + H\sum_{A,A'=2,5,7}\Big[|\psi^TC\lambda_f^A\lambda_c^{A'}\psi|^2 + |\psi^TC\gamma_5\lambda_f^A\lambda_c^{A'}\psi|^2 \Big] \ . \nonumber\\ \label{LFour}
\end{eqnarray}
In this Lagrangian, $\lambda_f^A$ and $\lambda_c^{A'}$ are the Gell-Mann matrices for  flavor and color spaces, respectively, and $C=i\gamma^2\gamma^0$ is the charge-conjugation Dirac matrix. As for the $H$ term in Eq.~(\ref{LFour}), we have included only  $A,A'=2,5,7$ channels which are antisymmetric with respect to both the flavor and color indices, since these combinations can generate the most attractive forces in between the two quarks~\cite{Buballa:2003qv}. At first glance, symmetric properties of the two terms in Eq.~(\ref{LFour}) are obscure because they are written in terms of parity-eigenstate bases. In order to see the properties more clearly, we introduce the following quark bilinear fields:
\begin{eqnarray}
\phi_{ij} &=& (\bar{\psi}_{R})^a_j(\psi_{L})^a_i \ ,\nonumber\\
(\eta_L)_i^a &=& \epsilon_{ijk}\epsilon^{abc}(\psi_L^T)_j^bC(\psi_L)_k^c\ , \nonumber\\
(\eta_R)_i^a &=& \epsilon_{ijk}\epsilon^{abc}(\psi_R^T)_j^bC(\psi_R)_k^c\ . \label{Bilinears}
\end{eqnarray}
In these fields, $\psi_{R(L)}=\frac{1\pm\gamma_5}{2}\psi$ is the right-handed (left-handed) quark, and the subscript ``$i,j,\cdots$'' and superscript ``$a,b,\cdots$'' represent flavor and color fundamental indices, respectively. Under $U(3)_L\times U(3)_R$ chiral transformation, $\psi_L$ and $\psi_R$ transform as $\psi_L\to g_L\psi_L$ and $\psi_R\to g_R\psi_R$, where $g_L\in U(3)_L$ ($g_R\in U(3)_R$), and accordingly, chiral transformation laws of $\phi$, $\eta_L$ and $\eta_R$ read
\begin{eqnarray}
\phi \to g_L\phi g_R^\dagger \ , \ \ \eta_L \to \eta_L g_L^\dagger\ , \ \ \eta_R \to \eta_R g_R^\dagger\ .
\end{eqnarray}
Meanwhile, one can easily derive identities 
\begin{eqnarray}
&& {\rm tr}[\phi^\dagger\phi] = \frac{1}{8}\sum_{A=0}^8\Big[(\bar{\psi}\lambda_f^A\psi)^2+(\bar{\psi}i\gamma_5\lambda_f^a\psi)^2\Big] \ , \nonumber\\
&&\eta_L^T\eta_L^* + \eta_R^T\eta_R^* = \frac{1}{2}\sum_{A,A'=2,5,7}\Big[|\psi^TC\lambda_f^A\lambda_c^{A'}\psi|^2 \nonumber\\
&& \hspace{30mm} + |\psi^TC\gamma_5\lambda_f^A\lambda_c^{A'}\psi|^2 \Big] \ ,
\end{eqnarray}
and hence, the four-point interaction Lagrangian~(\ref{LFour}) is rewritten into
\begin{eqnarray}
 {\cal L}_{4q} = 8G{\rm tr}[\phi^\dagger\phi] + 2H(\eta_L^T\eta_L^* + \eta_R^T\eta_R^*)\ . \label{LFourChiral}
\end{eqnarray}
Equation~(\ref{LFourChiral}) clearly shows $U(3)_L\times U(3)_R$ chiral symmetry of the interactions. Moreover, (global) $SU(3)_c$ color symmetry is also manifest since $\eta_L$ and $\eta_R$ belong to $\bar{\bm 3}$ representation of $SU(3)_c$ color group.

The third part of Eq.~(\ref{LNJL}), ${\cal L}_{6q}^{\rm anom.}$, is of the form
\begin{eqnarray}
{\cal L}_{6q}^{\rm anom.} &=& -8K({\rm det}\phi + {\rm det}\phi^\dagger)  \nonumber\\
&+& K'(\eta^T_L\phi\eta_R^*+\eta^T_R \phi^\dagger\eta_L^*)\ . \label{LSix}
\end{eqnarray}
This Lagrangian is again left unchanged under $SU(3)_L\times SU(3)_R$ chiral and $SU(3)_c$ color transformations but is not invariant under $U(1)_A$ axial transformation. That is, Eq.~(\ref{LSix}) is responsible for leading contributions from the $U(1)_A$ axial anomaly. In particular, $K$ term is often referred to as the Kobayashi-Maskawa-'t Hooft (KMT) determinant term which plays an essential role in generating a large mass of $\eta'$ meson~\cite{Kobayashi:1970ji,Kobayashi:1971qz,tHooft:1976snw,tHooft:1976rip}. The $K'$ term captures direct contributions of the anomaly effects to diquarks through interactions with mesons~\cite{Abuki:2010jq}. 

When we derive the four-point interactions in Eq.~(\ref{LFour}) from a one-gluon exchange vertex, the Fierz transformation yields $H/G=3/4$. Similarly, one obtain $K'/K=1/1$ from the Fierz transformation of the instanton vertex for the anomalous six-point interactions in Eq.~(\ref{LSix}). However, in the present study, we regard all of them as free parameters based on our effective-model description.

In the vacuum, i.e., at zero temperature and density, it is well known that (approximate) $SU(3)_L\times SU(3)_R$ chiral symmetry is spontaneously broken by emergence of chiral condensates such that quarks acquire their dynamical masses. Under an assumption of $SU(2)_I$ isospin symmetry, the chiral condensates are described by vacuum expectation values (VEVs) of $\phi$ as $\langle \phi\rangle = \frac{1}{2}{\rm diag}(\langle\bar{q}q\rangle,\langle\bar{q}q\rangle,\langle\bar{s}s\rangle)$. In cold and dense regime, VEVs of $\eta_L$ and $\eta_R$ can also become nonzero which represents emergence of the color superconducting phase~\cite{Buballa:2003qv,Alford:2007xm}. However, in our present study, we mainly focus on hot medium so that we do not take into account such condensates.

In order to evaluate fluctuations of diquark modes at the one-loop level of quarks, we make use of the following linearization of multi-quark couplings with the help of the Wick's theorem:
\begin{eqnarray}
&&XY \to XY+ \langle X\rangle Y+\langle Y\rangle X-\langle X\rangle\langle Y\rangle \ , \nonumber\\
&& XYZ \to \langle X\rangle YZ + \langle Y\rangle XZ + \langle Z\rangle XY + \langle X\rangle\langle Y\rangle Z \nonumber\\
&&\hspace{10mm} +\langle Y\rangle\langle Z\rangle X + \langle Z\rangle\langle X\rangle Y-2\langle X\rangle\langle Y\rangle\langle Z\rangle \ . \label{Approximation}
\end{eqnarray}
Within this approximation, from NJL Lagrangian~(\ref{LNJL}) dynamical masses of $q(=u,d)$ and $s$ quarks incorporating the chiral condensates are read off as
 \begin{eqnarray}
M_q &=& m_q-4G\langle\bar{q}q\rangle + 2K\langle\bar{q}q\rangle\langle\bar{s}s\rangle\ , \nonumber\\
M_s &=& m_s -4G\langle\bar{s}s\rangle + 2K\langle\bar{q}q\rangle^2\ , \label{ConsMass}
\end{eqnarray}
respectively. The mass fourmula~(\ref{ConsMass}) indicates that the nonanomalous $G$ term generates $\langle\bar{q}q\rangle$ ($\langle\bar{s}s\rangle$) contributions to $M_q$ ($M_s$). Meanwhile, the anomalous $K$ term generates $\langle\bar{s}s\rangle$ ($\langle\bar{q}q\rangle$) contributions to $M_q$ ($M_s$), that is, the latter term mixes different flavor contents. Besides, Eq.~(\ref{ConsMass}) indicates that the modification of $M_q$ is strongly controlled by the change of $\langle\bar{q}q\rangle$ regardless of the value of $K$ since $\langle\bar{q}q\rangle$ appears in both the $G$ and $K$ terms, while that of $M_s$ is not. These noteworthy features play important roles for temperature dependences of $M_q$ and $M_s$, as will be explained in Sec.~\ref{sec:Inputs}.

\section{Fluctuations of diquarks}
\label{sec:Analysis}

Diquark masses within the NJL model can be evaluated by pole positions of the corresponding Bethe-Salpeter (BS) amplitudes. In this section, we show our strategy to compute the BS amplitude.

\subsection{BS amplitude}
\label{sec:BSEquation}

 In this subsection, we provide explanations how to evaluate pole positions in the BS amplitudes for diquark channels~\cite{Hellstern:1997nv}.

The BS amplitude ${\cal T}$ is evaluated by infinite scatterings of the quarks as
\begin{eqnarray}
{\cal T} &=& {\cal K} + {\cal K}{\cal J}{\cal K} +  {\cal K}{\cal J}{\cal K} {\cal J}{\cal K}  + \cdots \nonumber\\
&=& {\cal K} + {\cal K}{\cal J}{\cal T}\ , 
\end{eqnarray}
namely,
\begin{eqnarray}
{\cal T} = (1-{\cal K}{\cal J})^{-1}{\cal K} \ , \label{TBSPole}
\end{eqnarray} 
where ${\cal K}$ is the interaction kernel and ${\cal J}$ is a loop function generated by quarks. In our present analysis, the kernels are read off from effective four-point interactions of quarks in Eq.~(\ref{LNJL}) with the approximation~(\ref{Approximation}). When we take parity eigenstates of the diquarks as
\begin{eqnarray}
(\eta_+)_i^a &=& \frac{1}{\sqrt{2}}(\eta_R-\eta_L)_i^a = \frac{1}{\sqrt{2}}\epsilon_{ijk}\epsilon^{abc}\psi_j^{T,b} C\gamma_5 \psi_k^c\ , \nonumber\\
(\eta_-)_i^a &=& \frac{1}{\sqrt{2}}(\eta_R+\eta_L)_i^a = \frac{1}{\sqrt{2}}\epsilon_{ijk}\epsilon^{abc}\psi_j^{T,b} C \psi_k^c\ , \label{Eta+-}
\end{eqnarray}
the kernel ${\cal K}$'s for respective diquark channels read
\begin{eqnarray}
{\cal K}_{[qq]_+}^{ab} &=& i\left(2H -\frac{K'}{2}\langle\bar{s}s\rangle\right)\delta^{ab} \ , \nonumber\\
{\cal K}_{[sq]_+}^{ab} &=& i\left(2H -\frac{K'}{2}\langle\bar{q}q\rangle\right)\delta^{ab} \ , \nonumber\\
{\cal K}_{[qq]_-}^{ab} &=& i\left(2H + \frac{K'}{2}\langle\bar{s}s\rangle\right)\delta^{ab}\ , \nonumber\\
{\cal K}_{[sq]_-}^{ab} &=& i\left(2H +\frac{K'}{2}\langle\bar{q}q\rangle\right)\delta^{ab} \ .\label{KernelD}
\end{eqnarray}
In these expressions, the subscripts $[qq]_\pm$ and $[sq]_\pm$ represent $(\eta_\pm)_{i=3}$ and $(\eta_\pm)_{i=1,2}$ channels, respectively, where the former is made of $u$ and $d$ quarks while the latter is of $u$ and $s$ (or $d$ and $s$) quarks. The color factor $\delta^{ab}$ guarantees the conservation of color charges. One sees that the $K'$ term incorporates the chiral condensates so as to induce the mass differences
of the positive- and negative-parity diquarks.

As for the loop functions, within the present quark one-loop approximation, ${\cal J}$'s for scalar diquark channels are given by
\begin{eqnarray}
{\cal J}_{[qq]_+}^{ab}(q) &=& 4i\delta^{ab}T\sum_m\int\frac{d^3p}{(2\pi)^3}{\rm tr}\Big[\gamma_5S_{(q)}(p')\gamma_5{S}^c_{(q)}(p)\Big]\ , \nonumber\\
{\cal J}_{[sq]_+}^{ab}(q) &=& 2i\delta^{ab}T\sum_m\int\frac{d^3p}{(2\pi)^3}{\rm tr}\Big[\gamma_5S_{(s)}(p')\gamma_5{S}^c_{(q)}(p) \nonumber\\
 && + \gamma_5S_{(q)}(p')\gamma_5{S}^c_{(s)}(p) \Big]\ , \label{JSDiquark}
\end{eqnarray}
and those for pseudoscalar diquark channels are by
\begin{eqnarray}
{\cal J}_{[qq]_-}^{ab}(q) &=& 4i\delta^{ab}T\sum_m\int\frac{d^3p}{(2\pi)^3}{\rm tr}\Big[S_{(q)}(p'){S}^c_{(q)}(p)\Big]\ , \nonumber\\
{\cal J}_{[sq]_-}^{ab}(q) &=& 2i\delta^{ab}T\sum_m\int\frac{d^3p}{(2\pi)^3}{\rm tr}\Big[S_{(s)}(p'){S}^c_{(q)}(p)  \nonumber\\
&& + S_{(q)}(p'){S}^c_{(s)}(p) \Big]\ , \label{JPSDiquark}
\end{eqnarray}
with $p'=p+q$. In these equations, we have defined propagators of $q$ quark by
\begin{eqnarray}
S_{(q)}(p) &=& {\rm F.T.}\langle 0|{\rm T} q(x)\bar{q}(0)|0\rangle= i\sum_{\zeta={\rm p}, {\rm a}}\frac{\Lambda^{(q)}_\zeta({\bm p})}{p_0-\eta_\zeta\epsilon^{(q)}_\zeta({\bm p})}\ , \nonumber\\
S^c_{(q)}(p) &=& {\rm F.T.}\langle 0|{\rm T} q^c(x)\bar{q}^c(0)|0\rangle= i\sum_{\zeta={\rm p}, {\rm a}}\frac{\Lambda_\zeta^{(q),c}({\bm p})}{p_0+\eta_\zeta\epsilon^{(q)}_\zeta({\bm p})} \ .\nonumber\\ \label{Propagator}
\end{eqnarray}
In Eq.~(\ref{Propagator}), we have introduced a propagator of the charge-conjugated quark field $q^c = C\bar{q}^T$ so as to perform the Dirac trace of the one loops straightforwardly. Besides, $\Lambda_\zeta^{(q)}$ is a projection operator with respect to the positive-energy and negative-energy contributions of the $q$ quark
\begin{eqnarray}
\Lambda^{(q)}_\zeta({\bm p}) = \frac{E_{\bm p}^{(q)}\gamma_0+\eta_\zeta(M_q-{\bm p}\cdot{\bm \gamma})}{2E^{(q)}_{\bm p}}\ , 
\end{eqnarray}
with $E_{\bm p}^{(q)} = \sqrt{{\bm p}^2+M_q^2}$ and $\eta_{\rm p}=+1$ ($\eta_{\rm a}=-1$). The charge-conjugated projection operator is simply given by interchanging the subscripts ${\rm p}$ and ${\rm a}$ as
\begin{eqnarray}
\Lambda^{(q),c}_{\rm p}({\bm p}) = \Lambda^{(q)}_{\rm a}({\bm p}) \ , \ \ \Lambda^{(q),c}_{\rm a}({\bm p}) = \Lambda^{(q)}_{\rm p}({\bm p})\ .
\end{eqnarray}
The single-quark dispersion relations affected by the chemical potential $\mu$ read
\begin{eqnarray}
\epsilon_\zeta^{(q)}({\bm p}) &=& E_{\bm p}^{(q)}-\eta_\zeta\mu \ .
\end{eqnarray}
In the same way, propagators of the $s$ quark are defined. It should be noted that, in Eqs.~(\ref{JSDiquark}) and~(\ref{JPSDiquark}), we have replaced $p_0$ integrations by the Matsubara summations $iT\sum_m$ with $p_0=i\omega_m\equiv i(2m+1)\pi T$ ($m\in \mathbb{Z}$) being the Matsubara frequencies in order to access hot matter~\cite{kapusta2006finite}.

As for the one-loop functions in Eqs.~(\ref{JSDiquark}) and~(\ref{JPSDiquark}), kinetic contributions stemming from spin couplings among quarks and diquarks are evaluated by the Dirac trace formulas ($f,f'=q\, {\rm or}\, s$)
\begin{eqnarray}
&&{\rm tr}[\Lambda^{(f')}_{\zeta'}({\bm p}')\Lambda_\zeta^{c,(f)}({\bm p})]  = 1+\frac{ \eta_{\zeta'}\eta_\zeta({\bm p}'\cdot{\bm p}-M_{f'}M_f)}{E^{(f')}_{{\bm p}'}E^{(f)}_{\bm p}} \ ,\nonumber\\
&&{\rm tr}[\gamma_5\Lambda^{(f')}_{\zeta'}({\bm p}')\gamma_5\Lambda_\zeta^{c,(f)}({\bm p})]  = -1 \nonumber\\
&& \hspace{40mm} -\frac{ \eta_{\zeta'}\eta_\zeta({\bm p}'\cdot{\bm p}+M_{f'}M_f)}{E^{(f')}_{{\bm p}'}E^{(f)}_{\bm p}}\ .\nonumber\\ \label{DiracTrace}
\end{eqnarray}
In addition, medium effects describing occupation probabilities of the quarks are incorporated by the Matsubara summation formula~\cite{kapusta2006finite}
\begin{eqnarray}
T\sum_m\frac{1}{(p_0'-\epsilon'_{{\bm p}'})(p_0-\epsilon_{\bm p})} = \frac{f_F(\epsilon_{\bm p})-f_F(\epsilon'_{{\bm p}'})}{i\bar{\omega}_n-\epsilon'_{{\bm p}'}+\epsilon_{\bm p}}\ ,  \nonumber\\
\label{MatsubaraSum}
\end{eqnarray}
with $\bar{\omega}_n \equiv 2n\pi T$ ($n\in\mathbb{Z}$), where $f_F(\epsilon)=1/({\rm e}^{\epsilon/T}+1)$ is the Fermi-Dirac distribution function. The retarded one-loop functions in our real-time world are obtained by the analytic continuation of $i\bar{\omega}_n\to q_0+i0$ in the denominator of Eq.~(\ref{MatsubaraSum}).

With the help of Eqs.~(\ref{DiracTrace}) and~(\ref{MatsubaraSum}), the one-loop functions in Eqs.~(\ref{JSDiquark}) and~(\ref{JPSDiquark}) can be evaluated.

\subsection{Regularization}
\label{sec:Regularization}

 In making use of the formula~(\ref{MatsubaraSum}), one encounters ultraviolet (UV) divergences when either $\epsilon_2({\bm p}')$ or $\epsilon_1({\bm p})$ is negative, which must be regularized. In the present study, we employ a three-dimensional proper-time regularization including not only a UV cutoff $\Lambda_{\rm UV}$ but also an infrared (IR) cutoff $\mu_{\rm IR}$~\cite{Ebert:1996vx,Hellstern:1997nv}. This treatment is implemented by the following replacement:
\begin{eqnarray}
\frac{1}{q_0-\epsilon'_{{\bm p}'}+\epsilon_{\bm p}+i0} \to \frac{{\cal R}(q_0-\epsilon'_{{\bm p}'}+\epsilon_{\bm p})}{q_0-\epsilon'_{{\bm p}'}+\epsilon_{\bm p}+i0}\ , \label{Regularization}
\end{eqnarray}
where the function ${\cal R}(x)$ is defined by
\begin{eqnarray}
{\cal R}(x) \equiv {\rm e}^{-\frac{|x|}{\Lambda_{\rm UV}}} - {\rm e}^{-\frac{|x|}{\mu_{\rm IR}}}\ . \label{RDef}
\end{eqnarray}
In this regularization, real and imaginary parts of Eq.~(\ref{Regularization}) are read off by the Cauchy principal-value integral
\begin{eqnarray}
 \frac{{\cal R}(q_0-\epsilon'_{{\bm p}'}+\epsilon_{\bm p})}{q_0-\epsilon'_{{\bm p}'}+\epsilon_{\bm p}+i0} &=&{\rm P} \frac{{\cal R}(q_0-\epsilon'_{{\bm p}'}+\epsilon_{\bm p})}{q_0-\epsilon'_{{\bm p}'}+\epsilon_{\bm p}} -i\pi \times \nonumber\\
 && \delta(q_0-\epsilon'_{{\bm p}'}+\epsilon_{\bm p}){\cal R}(q_0-\epsilon'_{{\bm p}'}+\epsilon_{\bm p})\ . \nonumber\\
\end{eqnarray}
Thus, the remaining ${\bm p}$ integration yields a relation $q_0-\epsilon'_{{\bm p}'}+\epsilon_{\bm p}=0$ from the delta function for the imaginary part, and using a property ${\cal R}(0)={\rm e}^0-{\rm e}^0=0$ one can find
\begin{eqnarray}
\frac{{\cal R}(q_0-\epsilon'_{{\bm p}'}+\epsilon_{\bm p})}{q_0-\epsilon'_{{\bm p}'}+\epsilon_{\bm p}+i0}  \overset{\rm Im}{=} 0\ . \label{RegImaginary}
\end{eqnarray}

Equation~(\ref{RegImaginary}) implies that all imaginary parts of the loop function ${\cal J}$'s are removed when both the UV and IR cutoffs $\Lambda_{\rm UV}$ and $\mu_{\rm IR}$ are introduced~\cite{Ebert:1996vx,Hellstern:1997nv}. Therefore, when focusing on diquarks in medium, the regularization~(\ref{Regularization}) allows us to remove physical processes of the Landau dampings such as $q\to [qq]+\bar{q}$ as well as those of the pair creations (annihilations) such as $[qq]\to q+q$ ($q+q\to [qq]$). Hence, the present regularization method enables us to define the masses of diquarks properly with no influence from imaginary parts. Intuitively speaking, disappearance of the physical processes implies that diquarks are doped into {\it color-singlet} matter and cannot discriminate colorful quarks in medium, while they indeed feel the medium effects. In this sense, diquarks are ``confined'' in our present study.\footnote{It was shown that not only imaginary parts but also poles of the BS amplitudes in the physical region disappear, when including the next-to-leading order of the rainbow approximation. In Ref.~\cite{Hellstern:1997nv}, the authors used the term ``confinement'' from these two observations.} Therefore, our present treatment is expected to be useful for delineating modifications of SHBs and $T_{cc}$ in medium where the diquarks are well confined in the hadrons. The existence of the IR cutoff in QCD is suggested by, e.g., the usefulness of inclusion of a bare gluon mass which represents nonperturbative nature of QCD both in the vacuum~\cite{Tissier:2011ey} and in medium~\cite{Suenaga:2019jjv,Kojo:2021knn}.\footnote{The massive-gluon description for chiral dynamics in low-energy QCD was discussed in, e.g., Refs.~\cite{Pelaez:2021tpq,Pelaez:2022rwx}.} We note that the disappearance of imaginary parts shown in Eq.~(\ref{RegImaginary}) follows a cancellation of UV and IR regulator parts in Eq.~(\ref{RDef}), so that the imaginary parts are left finite when the IR cutoff is not included. We also note that ${\cal R}(x) \to1$ is recovered when we take $\Lambda_{\rm UV}\to\infty$ and $\mu_{\rm IR}\to0$.

\subsection{Loop functions}
\label{sec:Loops}

We are now ready to get rather concrete analytic expressions of the loop functions. That is, using Eqs.~(\ref{DiracTrace}) and~(\ref{MatsubaraSum}) together with the regularization~(\ref{Regularization}), the loop function ${\cal J}$'s of the diquarks at rest ${\bm q}={\bm 0}$ read
\begin{widetext}
\begin{eqnarray}
{\cal J}^{ab}_{[qq]_+}(q_0) &=& 4i\delta^{ab} \int\frac{d^3p}{(2\pi)^3} \Bigg\{ T_{\rm pp}^{[qq]_+}({\bm p}) \frac{{\cal R}\big(q_0-2\epsilon_{\rm p}^{(q)}({\bm p})\big)}{q_0-2\epsilon_{\rm p}^{(q)}({\bm p})}\left[1-2f_F\left(\epsilon_{\rm p}^{(q)}({\bm p})\right)\right] \nonumber\\
&-& T_{\rm aa}^{[qq]_+}({\bm p}) \frac{{\cal R}\big(q_0+2\epsilon_{\rm a}^{(q)}({\bm p})\big)}{q_0+2\epsilon^{(q)}_{\rm a}({\bm p})}
\left[1-2f_F\left(\epsilon_{\rm a}^{(q)}({\bm p})\right)\right]  \Bigg\}\ ,\label{JUD+}
\end{eqnarray}
\begin{eqnarray}
{\cal J}_{[sq]_+}^{ab}(q_0) &=& 4i\delta^{ab}\int\frac{d^3p}{(2\pi)^3}  \Bigg\{ T_{\rm pp}^{[sq]_+}({\bm p}) \frac{{\cal R}\big(q_0-\epsilon_{\rm p}^{(s)}({\bm p})-\epsilon^{(q)}_{\rm p}({\bm p})\big)}{q_0-\epsilon_{\rm p}^{(s)}({\bm p})-\epsilon^{(q)}_{\rm p}({\bm p})}\left[1-f_F\left(\epsilon_{\rm p}^{(q)}({\bm p})\right)-f_F\left(\epsilon^{(s)}_{\rm p}({\bm p})\right)\right] \nonumber\\
&+&T_{\rm pa}^{[sq]_+}({\bm p})\frac{{\cal R}\big(q_0-\epsilon_{\rm p}^{(s)}({\bm p})+\epsilon^{(q)}_{\rm a}({\bm p})\big)}{q_0-\epsilon_{\rm p}^{(s)}({\bm p})+\epsilon^{(q)}_{\rm a}({\bm p})}\left[f_F\left(\epsilon_{\rm a}^{(q)}({\bm p})\right)-f_F\left(\epsilon^{(s)}_{\rm p}({\bm p})\right)\right]  \nonumber\\
&-&T_{\rm ap}^{[sq]_+}({\bm p})\frac{{\cal R}\big(q_0+\epsilon_{\rm a}^{(s)}({\bm p})-\epsilon^{(q)}_{\rm p}({\bm p})\big)}{q_0+\epsilon_{\rm a}^{(s)}({\bm p})-\epsilon^{(q)}_{\rm p}({\bm p})}\left[f_F\left(\epsilon_{\rm p}^{(q)}({\bm p})\right)-f_F\left(\epsilon^{(s)}_{\rm a}({\bm p})\right)\right]  \nonumber\\
&-&T_{\rm aa}^{[sq]_+}({\bm p})\frac{{\cal R}\big(q_0+\epsilon_{\rm a}^{(s)}({\bm p})+\epsilon^{(q)}_{\rm a}({\bm p})\big)}{q_0+\epsilon_{\rm a}^{(s)}({\bm p})+\epsilon^{(q)}_{\rm a}({\bm p})}\left[1-f_F\left(\epsilon_{\rm a}^{(q)}({\bm p})\right)-f_F\left(\epsilon^{(s)}_{\rm a}({\bm p})\right)\right]  \Bigg\}  \ , \label{JSU+}
\end{eqnarray}
\begin{eqnarray}
{\cal J}^{ab}_{[qq]_-} (q_0)&=& 4i\delta^{ab}\int\frac{d^3p}{(2\pi)^3}\Bigg\{ T_{\rm pp}^{[qq]_-}({\bm p})\frac{{\cal R}\big(q_0-2\epsilon_{\rm p}^{(q)}({\bm p})\big)}{q_0-2\epsilon_{\rm p}^{(q)}({\bm p})}\left[1-2f_F\left(\epsilon_{\rm p}^{(q)}({\bm p})\right)\right] \nonumber\\
&+&2T_{\rm pa}^{[qq]_-}({\bm p})\frac{{\cal R}\big(q_0-\epsilon_{\rm p}^{(q)}({\bm p})+\epsilon^{(q)}_{\rm a}({\bm p})\big)}{q_0-\epsilon_{\rm p}^{(q)}({\bm p})+\epsilon^{(q)}_{\rm a}({\bm p})}\left[f_F\left(\epsilon_{\rm a}^{(q)}({\bm p})\right)-f_F\left(\epsilon^{(q)}_{\rm p}({\bm p})\right)\right] \nonumber\\
&-&T_{\rm aa}^{[qq]_-}({\bm p})\frac{{\cal R}\big(q_0+2\epsilon^{(q)}_{\rm a}({\bm p})\big)}{q_0+2\epsilon^{(q)}_{\rm a}({\bm p})}
\left[1-2f_F\left(\epsilon_{\rm a}^{(q)}({\bm p})\right)\right]  \Bigg\} \ ,\label{JUD-}
\end{eqnarray}
and
\begin{eqnarray}
{\cal J}^{ab}_{[sq]_-}(q_0) &=& 4i\delta^{ab}\int\frac{d^3p}{(2\pi)^3}\Bigg\{T_{\rm pp}^{[sq]_-}({\bm p})\frac{{\cal R}\big(q_0-\epsilon_{\rm p}^{(s)}({\bm p})-\epsilon^{(q)}_{\rm p}({\bm p})\big)}{q_0-\epsilon_{\rm p}^{(s)}({\bm p})-\epsilon^{(q)}_{\rm p}({\bm p})}\left[1-f_F\left(\epsilon_{\rm p}^{(q)}({\bm p})\right)-f_F\left(\epsilon^{(s)}_{\rm p}({\bm p})\right)\right] \nonumber\\
&+&T_{\rm pa}^{[sq]_-}({\bm p})\frac{{\cal R}\big(q_0-\epsilon_{\rm p}^{(s)}({\bm p})+\epsilon^{(q)}_{\rm a}({\bm p})\big)}{q_0-\epsilon_{\rm p}^{(s)}({\bm p})+\epsilon^{(q)}_{\rm a}({\bm p})}\left[f_F\left(\epsilon_{\rm a}^{(q)}({\bm p})\right)-f_F\left(\epsilon^{(s)}_{\rm p}({\bm p})\right)\right] \nonumber\\
&-& T_{\rm ap}^{[sq]_-}({\bm p})\frac{{\cal R}\big(q_0+\epsilon_{\rm a}^{(s)}({\bm p})-\epsilon^{(q)}_{\rm p}({\bm p})\big)}{q_0+\epsilon_{\rm a}^{(s)}({\bm p})-\epsilon^{(q)}_{\rm p}({\bm p})}\left[f_F\left(\epsilon_{\rm p}^{(q)}({\bm p})\right)-f_F\left(\epsilon^{(s)}_{\rm a}({\bm p})\right)\right] \nonumber\\
&-&T_{\rm aa}^{[sq]_-}({\bm p})\frac{{\cal R}\big(q_0+\epsilon^{(s)}_{\rm a}({\bm p})+\epsilon_{\rm a}^{(q)}({\bm p})\big)}{q_0+\epsilon^{(s)}_{\rm a}({\bm p})+\epsilon_{\rm a}^{(q)}({\bm p})}
\left[1-f_F\left(\epsilon_{\rm a}^{(q)}({\bm p})\right)-f_F\left(\epsilon^{(s)}_{\rm a}({\bm p})\right)\right]  \Bigg\} \ , \label{JSU-}
\end{eqnarray}
\end{widetext}
where we have defined the kinetic contributions stemming from spin-coupling properties by
\begin{eqnarray}
&&T^{[qq]_+}_{\rm pp}({\bm p}) = T^{[qq]_+}_{\rm aa}({\bm p}) =  2\ , \nonumber\\
&& T^{[sq]_+}_{\rm pp}({\bm p}) =  T^{[sq]_+}_{\rm aa}({\bm p}) = 1+\frac{{\bm p}^2+M_qM_s}{E_{\bm p}^{(q)}E_{\bm p}^{(s)}}\ , \nonumber\\
&& T^{[sq]_+}_{\rm pa}({\bm p}) =  T^{[sq]_+}_{\rm ap}({\bm p}) = 1-\frac{{\bm p}^2+M_qM_s}{E_{\bm p}^{(q)}E_{\bm p}^{(s)}}\ ,  \label{Kin+}
\end{eqnarray}
for positive-parity diquarks and
\begin{eqnarray}
&&T^{[qq]_-}_{\rm pp}({\bm p}) = T^{[qq]_-}_{\rm aa}({\bm p}) =  \frac{2{\bm p}^2}{\big(E_{\bm p}^{(q)}\big)^2}\ , \nonumber\\
&&T^{[qq]_-}_{\rm pa}({\bm p}) = \frac{2M_q^2}{\big(E_{\bm p}^{(q)}\big)^2}\ , \nonumber\\
&& T^{[sq]_-}_{\rm pp}({\bm p}) =  T^{[sq]_-}_{\rm aa}({\bm p}) = 1+\frac{{\bm p}^2-M_qM_s}{E_{\bm p}^{(q)}E_{\bm p}^{(s)}} \ , \nonumber\\
&& T^{[sq]_-}_{\rm pa}({\bm p}) =  T^{[sq]_-}_{\rm ap}({\bm p}) = 1-\frac{{\bm p}^2-M_qM_s}{E_{\bm p}^{(q)}E_{\bm p}^{(s)}} , \label{Kin-}
\end{eqnarray}
for negative-parity diquarks. The differences of these quantities between the chiral partners are proportional to $M_q$ so that they vanish in a limit of $M_q\to0$, e.g., $T^{[qq]_+}_{\rm pp}({\bm p}) = T^{[qq]_-}_{\rm pp}({\bm p})  \to2$ and $T^{[sq]_+}_{\rm pp}({\bm p}) = T^{[sq]_-}_{\rm pp}({\bm p})  \to 1 + |{\bm p}|/E_{\bm p}^{(q)}$, which reflects a fact that all diquarks examined in this work contain at least one $q$ quark. 

From Eqs.~(\ref{JUD+}) and~(\ref{JUD-}), for instance, one can see that variations of the loop functions between the chiral partners are solely incorporated by the kinetic contributions~(\ref{Kin+}) and~(\ref{Kin-}). This is easily understood since $[qq]_+$ and $[qq]_-$ diquarks differ by only their spin structures; the former is$\,^1S_0$ while the latter is$\,^3P_0$ in a quark-model sense, with the same color and flavor contents. Similar structures are found in $[sq]_+$ and $[sq]_-$ sectors from Eqs.~(\ref{JSU+}) and~(\ref{JSU-}).

The loop functions of antidiquarks $[\bar{q}\bar{q}]_\pm$ and $[\bar{s}\bar{q}]_\pm$ are simply evaluated by changing the variable $q_0$ to $-q_0$ in the corresponding ${\cal J}$'s as
\begin{eqnarray}
&& {\cal J}^{ab}_{[\bar{q}\bar{q}]_+}(q_0) = {\cal J}^{ab}_{[{q}q]_+}(-q_0)\ , \ \ {\cal J}^{ab}_{[\bar{s}\bar{q}]_+}(q_0) = {\cal J}^{ab}_{[{s}q]_+}(-q_0)\ , \nonumber\\
&& {\cal J}^{ab}_{[\bar{q}\bar{q}]_-}(q_0) = {\cal J}^{ab}_{[{q}q]_-}(-q_0)\ , \ \ {\cal J}^{ab}_{[\bar{s}\bar{q}]_-}(q_0) = {\cal J}^{ab}_{[{s}q]_-}(-q_0)\ , \nonumber\\
\end{eqnarray}
due to the charge-conjugation properties.

\section{Numerical results}
\label{sec:Results}

In this section, we present our numerical results of mass changes of the diquarks in medium.

\subsection{Inputs}
\label{sec:Inputs}

Before showing the results, in this subsection we explain our strategy to fix the model parameters.

Our NJL model contains eight parameters: the current quark masses $m_q$ and $m_s$, four-point couplings $G$ and $H$, six-point couplings responsible for the $U(1)_A$ axial anomaly $K$ and $K'$,  the UV cutoff $\Lambda_{\rm UV}$ and the IR cutoff $\mu_{\rm IR}$. Since $H$ and $K'$ do not affect pseudoscalar meson properties in our treatment, first we fix the remaining $m_q$, $m_s$, $G$, $K$, $\Lambda_{\rm UV}$ and $\mu_{\rm IR}$ from the meson sector.

As for inputs from the pseudoscalar sector, we adopt vacuum values of a pion mass, a kaon mass, a pion decay constant and a kaon decay constant as~\cite{Workman:2022ynf}
\begin{eqnarray}
&&m_\pi=0.138\, {\rm GeV}\ , \ \ m_K = 0.496\, {\rm GeV}\ ,  \nonumber\\
&&f_\pi=0.0921\, {\rm GeV}\ , \ \ f_K = 0.110\, {\rm GeV}\ . \label{InputVac}
\end{eqnarray} 
These inputs allow us to determine the four parameters $m_q$, $m_s$, $G$, $K$ for a given set of $\Lambda_{\rm UV}$ and $\mu_{\rm IR}$. The latter two parameters are fixed so as to derive a reasonable temperature dependence of the chiral condensates $\langle\bar{q}q\rangle$ and $\langle\bar{s}s\rangle$ at $\mu=0$, where lattice simulations already explored~\cite{Aoki:2009sc}. Evaluations of $m_\pi$, $m_K$, $f_\pi$ and $f_K$ in our present NJL model are not concise and not our main aim of the present work, so we leave them to Appendix~\ref{sec:PSInput}. The chiral condensates $\langle\bar{q}q\rangle$ and $\langle\bar{s}s\rangle$ at finite $T$ and $\mu$ are computed by
\begin{eqnarray}
\langle\bar{q}q\rangle &=& -3iT\sum_n\int\frac{d^3p}{(2\pi)^3}{\rm tr}[S_{(q)}(p)] \nonumber\\
&=& -12M_q\int\frac{d^3p}{(2\pi)^3}\frac{{\cal R}\big(\epsilon^{(q)}_{\rm p}({\bm p})+\epsilon^{(q)}_{\rm a}({\bm p})\big)}{\epsilon^{(q)}_{\rm p}({\bm p})+\epsilon^{(q)}_{\rm a}({\bm p})} \nonumber\\
&\times&\left[1-f_F\left(\epsilon_{\rm p}^{(q)}({\bm p})\right)-f_F\left(\epsilon_{\rm a}^{(q)}({\bm p})\right)\right]\ , \label{QQAnalytic}
\end{eqnarray}
and
\begin{eqnarray}
\langle\bar{s}s\rangle &=& -3iT\sum_n\int\frac{d^3p}{(2\pi)^3}{\rm tr}[S_{(s)}(p)]  \nonumber\\
&=& -12M_s\int\frac{d^3p}{(2\pi)^3}\frac{{\cal R}\big(\epsilon^{(s)}_{\rm p}({\bm p})+\epsilon^{(s)}_{\rm a}({\bm p})\big)}{\epsilon^{(s)}_{\rm p}({\bm p})+\epsilon^{(s)}_{\rm a}({\bm p})} \nonumber\\
&\times&\left[1-f_F\left(\epsilon_{\rm p}^{(s)}({\bm p})\right)-f_F\left(\epsilon_{\rm a}^{(s)}({\bm p})\right)\right]\ ,   \label{SSAnalytic}
\end{eqnarray}
respectively.

\begin{figure}[t]
\centering
\hspace*{-0.5cm} 
\includegraphics*[scale=0.8]{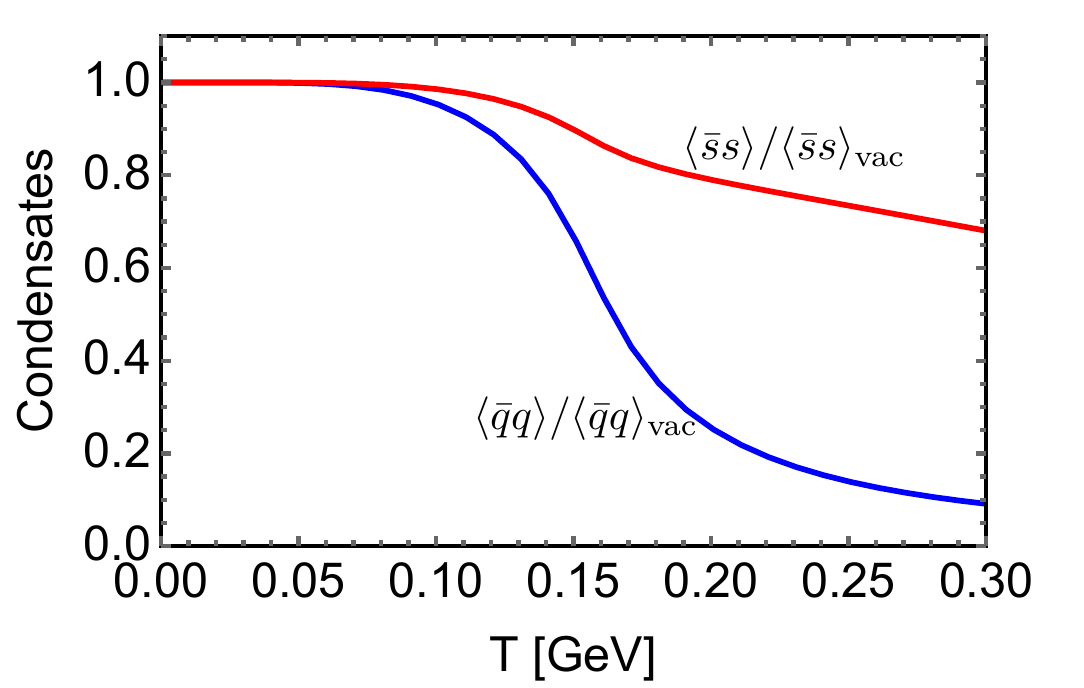}
\caption{Temperature dependence of the chiral condensates $\langle\bar{q}q\rangle$ and $\langle\bar{s}s\rangle$ at vanishing quark chemical potential $\mu=0$ normalized by their vacuum values.}
\label{fig:Condensates}
\end{figure}

\begin{figure*}[t]
\centering
\hspace*{-0.5cm} 
\includegraphics*[scale=0.62]{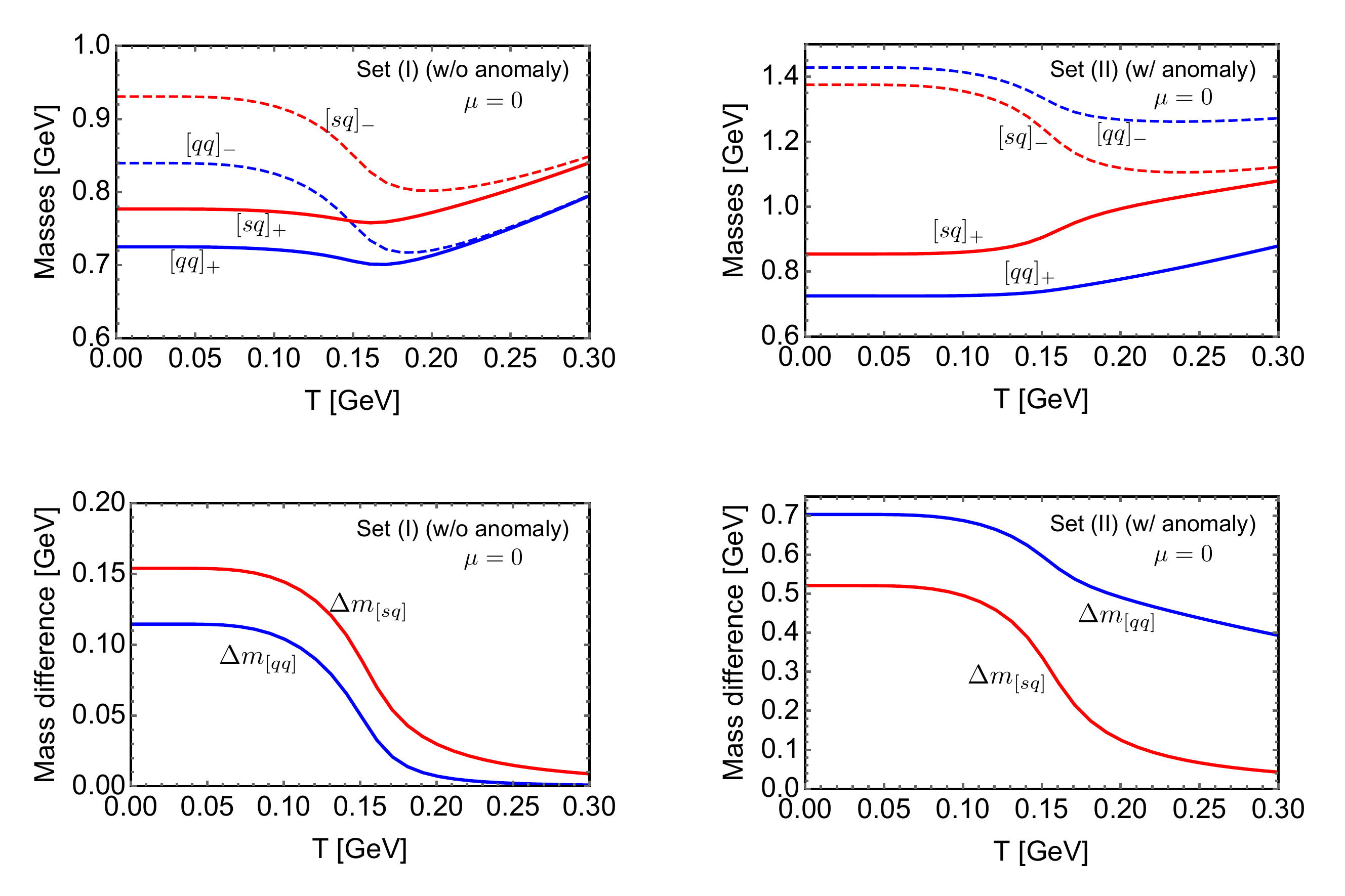}
\caption{Temperature dependence of the diquark masses for the parameter Set (I) (left) with no anomaly effects and Set (II) (right) with significant anomaly effects at $\mu=0$. The mass differences are defined by $\Delta m_{[qq]} =m_{[qq]_-}-m_{[qq]_+}$ and $\Delta m_{[sq]} =m_{[sq]_-}-m_{[sq]_+}$.}
\label{fig:DMass_Mu0}
\end{figure*}

When we take
\begin{eqnarray}
\Lambda_{\rm UV} = 1.6\, {\rm GeV}\ , \ \ \mu_{\rm IR} = 0.45\, {\rm GeV}\ , \label{Cutoffs}
\end{eqnarray}
and determine the model parameters
\begin{eqnarray}
&& m_q=0.00258\, {\rm GeV}\ , \ \ m_s = 0.0761\, {\rm GeV}\ , \nonumber\\
&& G=1.15\, {\rm GeV}^{-2}\ , \ \ K=10.3\, {\rm GeV}^{-5}\ , \label{Parameters}
\end{eqnarray}
from fitting the inputs~(\ref{InputVac}), the resultant $T$ dependence of the chiral condensates is obtained as depicted in Fig.~\ref{fig:Condensates}. Although reduction of $\langle\bar{s}s\rangle$ at finite temperature is slightly slow compared to lattice results, the pseudocritical temperature for $\langle\bar{q}q\rangle$ reads
\begin{eqnarray}
T_{\rm pc} \sim 0.15\,  {\rm GeV}\ , \label{TcChiral}
\end{eqnarray}
which is close to the lattice estimation~\cite{Aoki:2009sc}. 
Therefore, we expect that the parameters~(\ref{Cutoffs}) and~(\ref{Parameters}) are capable of capturing qualitative behavior of chiral symmetry in medium well, and in what follows we adopt those parameters.\footnote{We note that the mass of $\eta'$ meson cannot be reproduced by the parameter set~(\ref{Parameters}). However, our present study aims to shed light on the diquark masses in medium which are dominantly controlled by the remaining parameters $H$ and $K'$, and hence, we expect that the following discussion is not affected by this shortcoming considerably at a qualitative level. }

As seen from Eq.~(\ref{ConsMass}), $M_q$ is proportional to $\langle\bar{q}q\rangle$ when ignoring the current quark mass $m_q$ which is indeed small. Hence, at finite temperature $M_q$ drops substantially above $T_{\rm pc}\sim 0.15$ GeV in accordance with the sufficient reduction of $\langle\bar{q}q\rangle$ as in Fig.~\ref{fig:Condensates}. On the other hand, $M_s$ is generated by the comparably large $m_s$ and $\langle\bar{s}s\rangle$ contributions, so $M_s$ does not decrease prominently even above $T_{\rm pc}$. In Sec.~\ref{sec:DMassT}, we find that the former fast reduction of $M_q$ significantly affects the mass degeneracy of the chiral partners of diquarks above $T_{\rm pc}$.

One of our aims in this work is to examine the $U(1)_A$ axial anomaly effects to the chiral-partner structures of diquarks in medium. For this reason, as for the parameters in terms of diquarks, we regard $K'$ as a free parameter and determine the remaining $H$ from a lattice result: $m_{[qq]_+}^{\rm lattice}=0.725$ GeV~\cite{Bi:2015ifa}. In particular, we use two parameter sets with no anomaly effects ($K'=0$) and with significant effects ($K'=15$ GeV$^{-5}$). The determined $H$ and diquark masses in the vacuum with those values of $K'$ are tabulated in Table~\ref{tab:DiquarkMass}. It should be noted that we employ $K'=15$ GeV$^{-5}$ as a typical value of $K'$ such that the mass of chiral-partner $\Lambda_c(1/2^-)$ reads $M[\Lambda_c(1/2^-)]=2.99$ GeV. The detailed discussion in terms of the $\Lambda_c$ baryons is provided in Sec.~\ref{sec:Decays}.
\begin{table}[htbp]
\begin{center}
  \begin{tabular}{c||cc|cccc} \hline
& $K'$ & $H$ & $m_{[qq]_+}$ & $m_{[qq]_-}$ & $m_{[sq]_+}$ & $m_{[sq]_-}$ \\ 
& [GeV$^{-5}$] & [GeV$^{-2}$] & [GeV] & [GeV] & [GeV] & [GeV]\\  \hline 
Set(I) & $0$ & $1.77$  & $0.725^*$ & $0.840$ & $0.777$ & $0.931$ \\
Set(II) & $15$ & $1.53$  & $0.725^*$ & $1.43$ & $0.854$ & $1.38$   \\ \hline
 \end{tabular}
\caption{Determined values of $H$ and the diquark masses for two parameter sets with $K'=0$ and $K'=15$ GeV$^{-5}$. The asterisk ($*$) stands for an input from the lattice simulation.}
\label{tab:DiquarkMass}
\end{center}
\end{table}

From the table, one can see that, for positive-parity diquarks, $m_{[qq]_+}<m_{[sq]_+}$ follows as naively expected since the $s$ quark mass is larger than the $q$ quark mass. For negative-parity diquarks, $m_{[qq]_-}>m_{[sq]_-}$ which is often referred to as the inverse mass hierarchy is realized when significant anomaly effects enter~\cite{Harada:2019udr}, while the normal mass hierarchy $m_{[qq]_-}<m_{[sq]_-}$ is obtained when the anomaly effects are absent.

\begin{figure*}[t]
\centering
\hspace*{-0.5cm} 
\includegraphics*[scale=0.62]{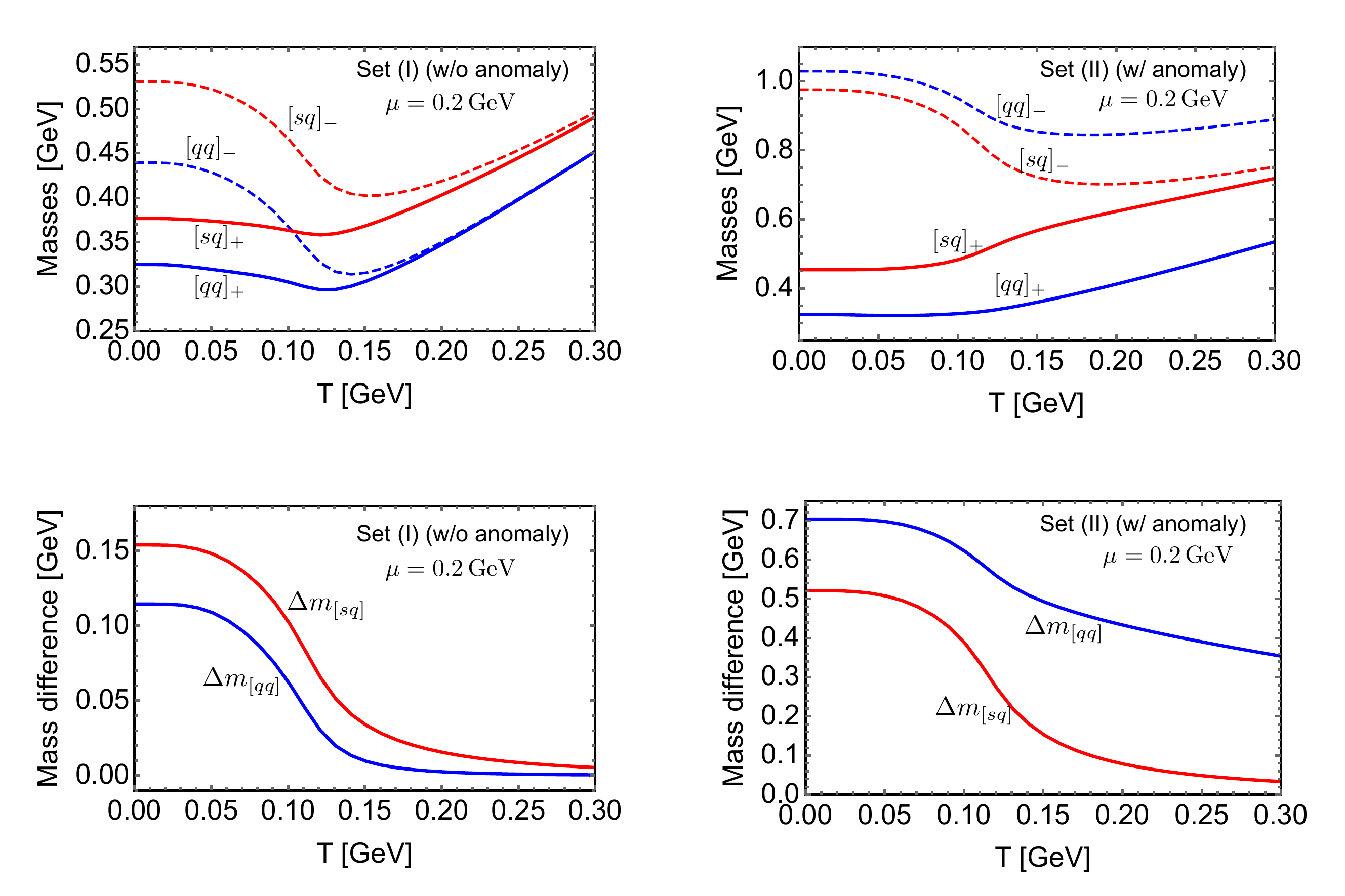}
\caption{Temperature dependence of the diquark masses and mass differences for the Set (I) (left) and Set (II) (right) at $\mu=0.2$ GeV.}
\label{fig:DMass_Mu200}
\end{figure*}
\begin{figure*}[htbp]
\centering
\hspace*{-0.5cm} 
\includegraphics*[scale=0.62]{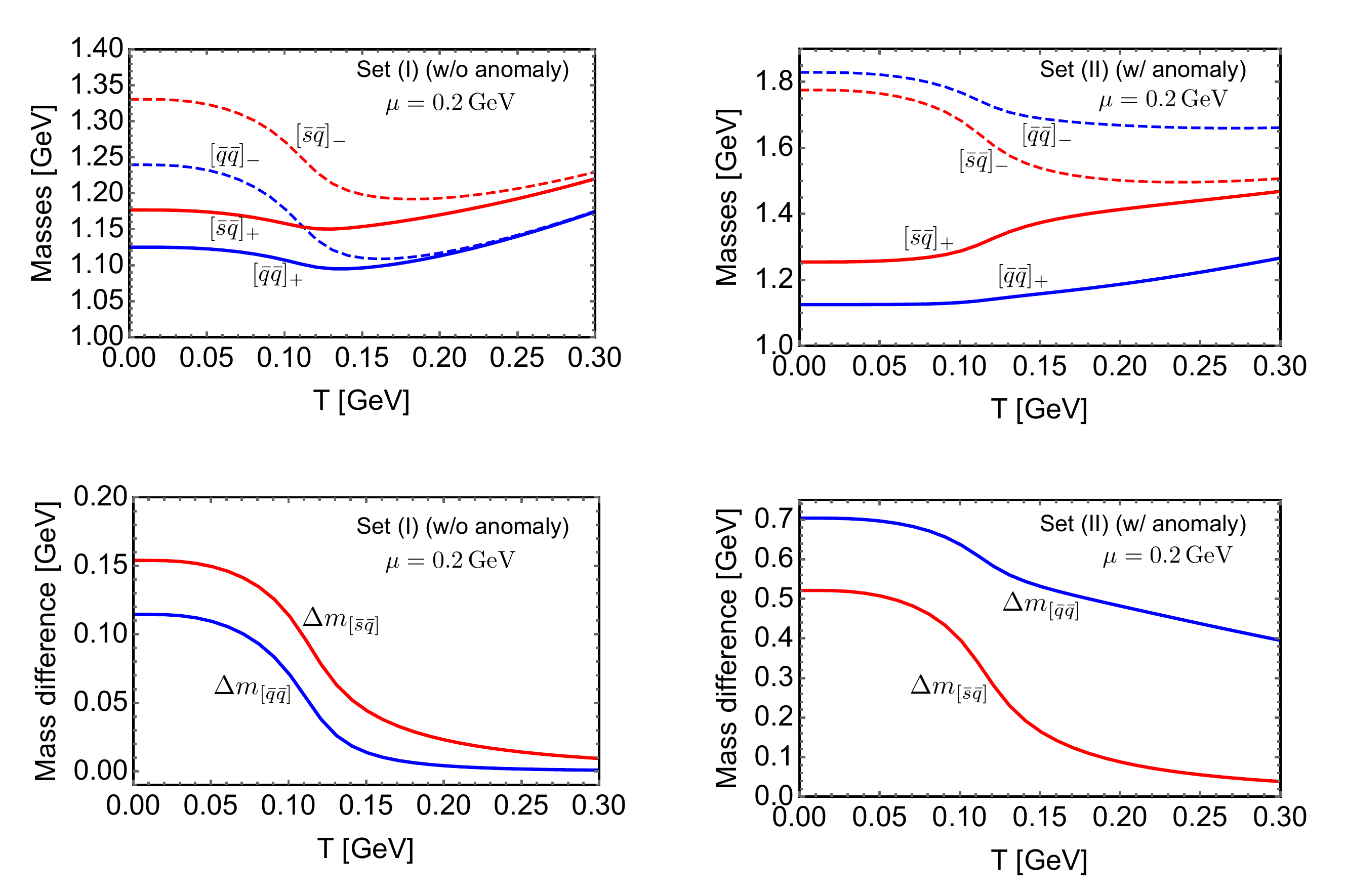}
\caption{Temperature dependence of the antidiquark masses and mass differences for the Set (I) (left) and Set (II) (right) at $\mu=0.2$ GeV. The mass differences are defined by $\Delta m_{[\bar{q}\bar{q}]} =m_{[\bar{q}\bar{q}]_-}-m_{[\bar{q}\bar{q}]_+}$ and $\Delta m_{[\bar{s}\bar{q}]} =m_{[\bar{s}\bar{q}]_-}-m_{[\bar{s}\bar{q}]_+}$.}
\label{fig:ADMass_Mu200}
\end{figure*}

\subsection{$T$ dependence of the diquark masses}
\label{sec:DMassT}

Here, we present our numerical results on temperature dependences of the diquark masses at a given $\mu$, and see their chiral-partner structures together with effects from the $U(1)_A$ axial anomaly. We note that, in the present paper, we define the diquark masses including the shifts from $\mu$.

Depicted in Fig.~\ref{fig:DMass_Mu0} is the resultant temperature dependences of the diquark masses at $\mu=0$ for the parameter Set (I) with no anomaly effects and Set (II) with significant anomaly effects in Table~\ref{tab:DiquarkMass}. Mass differences between the chiral partners, $\Delta m_{[qq]} \equiv m_{[qq]_-} - m_{[qq]_+}$ and $\Delta m_{[sq]} \equiv m_{[sq]_-} - m_{[sq]_+}$, are also displayed to see the chiral-partner structures more clearly. 

The left panels of Fig.~\ref{fig:DMass_Mu0} indicate that the normal mass hierarchy for both the positive and negative-parity diquarks is observed at any temperature when the anomaly effects are absent. Besides, the mass difference between the chiral partners always satisfies $\Delta m_{[sq]}>\Delta m_{[qq]}$. In the absence of the anomaly effects, the kernels for $[qq]_+$ and $[qq]_-$ ($[sq]_+$ and $[sq]_-$) are identical, and only the kinetic contributions in the loop functions~(\ref{Kin+}) and~(\ref{Kin-}) generate different effects to the two diquarks. As explained below Eqs.~(\ref{Kin+}) and~(\ref{Kin-}), such differences are proportional to $M_q$ which sufficiently drops above $T_{\rm pc}$ for both the $[qq]$ and $[sq]$ diquark sectors. Therefore, the mass degeneracy of the chiral partners takes place prominently above $T_{\rm pc}$, and as a result the chiral-partner structures are clearly seen at high temperature.

On the other hand, from the right panels of Fig.~\ref{fig:DMass_Mu0}, one can see that the inverse hierarchy for the negative-parity diquarks is always realized accompanied by the significant anomaly effects. Moreover, the mass difference reads $\Delta m_{[qq]}>\Delta m_{[sq]}$ at any temperature when the anomaly is switched on, and the reduction of $\Delta m_{[qq]}$ at high temperature is tempered whereas $\Delta m_{[sq]}$ is sufficiently suppressed similarly to with $K'=0$. The former tempered reduction is understood as follows. When $K'=15$ GeV$^{-5}$, the kernels ${\cal K}_{[qq]_+}$ and ${\cal K}_{[qq]_-}$ are significantly affected by the $\langle\bar{s}s\rangle$ contributions in addition to the constant $H$, signs of which are opposite for $[qq]_+$ and $[qq]_-$ channels as seen from Eq.~(\ref{KernelD}). Besides, the suppression of $\langle\bar{s}s\rangle$ at finite temperature is hindered compared to $\langle\bar{q}q\rangle$ as in Fig.~\ref{fig:Condensates}. Hence, the difference between ${\cal K}_{[qq]_+}$ and ${\cal K}_{[qq]_-}$ is left sizable even at $T\sim 0.3$ GeV and the resultant $\Delta m_{qq}$ also reads considerably large, although ${\cal J}_{[qq]_+}$ and ${\cal J}_{[qq]_-}$ become approximately identical. Meanwhile, the sufficient reduction of $\Delta m_{sq}$ is straightforwardly understood from the fast decrement of $\langle\bar{q}q\rangle$ above $T_{\rm pc}$, since the difference between ${\cal K}_{[sq]_+}$ and ${\cal K}_{[sq]_-}$ induced by $K'$ contributions is proportional to $\langle\bar{q}q\rangle$ and the situation is similar to with $K'=0$ at such high temperature. 

As for behaviors of the diquark masses at temperature, particularly from the top-left panel of Fig.~\ref{fig:DMass_Mu0}, one can see that $m_{[qq]_-}$ and $m_{[sq]_-}$ once decrease around $T_{\rm pc}$ but they turn to the increment above $T_{\rm pc}$. Such nonlinear temperature dependences are driven by the abrupt reduction of $\langle\bar{q}q\rangle$, namely, chiral-symmetry restoration. In fact, when we neglect effects from chiral-symmetry restoration, it can be shown that diquark masses exhibit monotonic increments as the system is heated owing to thermal effects~\cite{kapusta2006finite}. Reflecting such a thermal-mass property, at sufficiently high temperature where chiral symmetry is mostly restored the diquark masses monotonically increase. 

In order to see effects from a quark chemical potential $\mu$ to diquark masses at finite temperature, in Fig.~\ref{fig:DMass_Mu200} we also depict the resultant diquark masses and mass differences as a function of $T$ at $\mu=0.2$ GeV. The figure indicates that at $T=0$ all the diquark masses are reduced by $2\mu=0.4$ GeV since the diquarks carry the quark number $+2$. Besides, mass degeneracies of the chiral partners take place at a lower temperature compared to Fig.~\ref{fig:DMass_Mu0}, reflecting the fact that the pseudocritical temperature $T_{\rm pc}$ decreases at finite quark chemical potential. Except for these points, temperature dependences of the diquark masses are qualitatively similar to the ones at $\mu=0$.

At $\mu=0.2$ GeV, diquark and antiduquark masses differ due to the breakdown of charge-conjugation symmetry. Thus, in order to quantify such violation it is worth investigating temperature dependences of the antidiquark masses as well. The resultant temperature dependences are displayed in Fig.~\ref{fig:ADMass_Mu200}. In contrast to the results for the diquark masses in Fig.~\ref{fig:DMass_Mu200}, the masses of antidiquarks carrying the quark number $-2$ are increased by $2\mu=0.4$ GeV at $T=0$. Besides, at higher temperature where the mass degeneracies of the chiral partners occur substantially, increment of the antidiquark masses is tempered compared to those of the diquark masses. Those are major consequences of the violation of charge-conjugation symmetry. Meanwhile, the temperature dependences of mass differences between the partners: $\Delta m_{[\bar{q}\bar{q}]} \equiv m_{[\bar{q}\bar{q}]_-} -m_{[\bar{q}\bar{q}]_+}$ and $\Delta m_{[\bar{s}\bar{q}]} \equiv m_{[\bar{s}\bar{q}]_-} -m_{[\bar{s}\bar{q}]_+}$, are similar to those of $\Delta m_{[qq]}$ and $\Delta m_{[sq]}$, respectively.

\begin{figure}[htbp]
\centering
\hspace*{-0.5cm} 
\includegraphics*[scale=0.7]{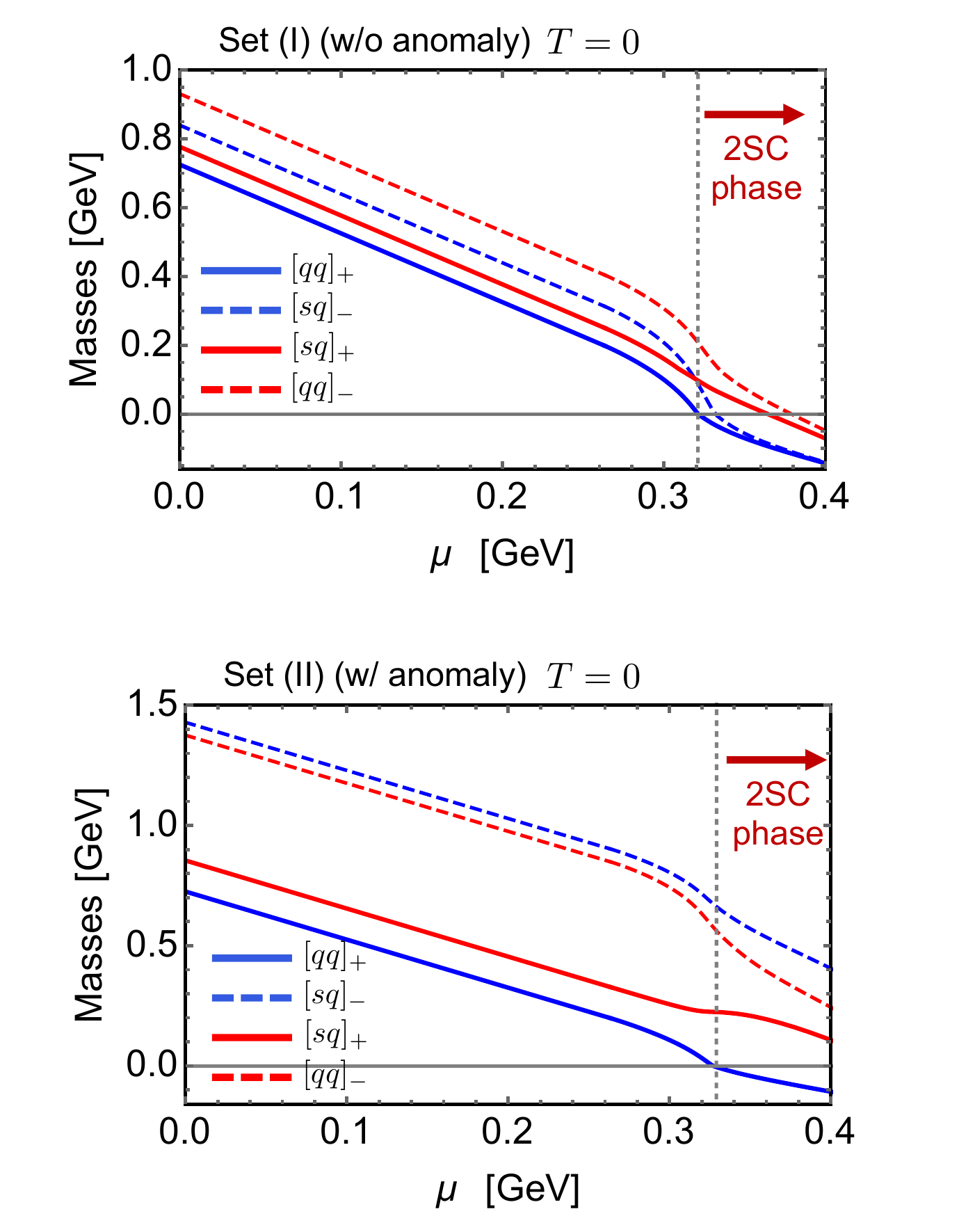}
\caption{Quark chemical potential dependence of the diquark masses for the Set (I) (top) and Set (II) (bottom) at $T=0$. The dotted vertical line corresponds to the critical chemical potential $\mu^*$ at which $m_{[qq]_+}=0$ is satisfied.}
\label{fig:DMass_T0}
\end{figure}

\subsection{$\mu$ dependence of the diquark masses}
\label{sec:DMassMu}

In Sec.~\ref{sec:DMassT}, the temperature dependences of the diquark masses have been examined and we have succeeded in getting deeper insights into the chiral-partner structures of the diquarks and $U(1)_A$ axial anomaly effects to them in hot matter. In this subsection, we investigate $\mu$ dependences of the diquark masses at $T=0$ in order to see the chiral-partner structures and anomaly effects in cold dense matter.

Depicted in Fig.~\ref{fig:DMass_T0} is the resultant $\mu$ dependences of the diquark masses for the parameter Set (I) (top) and Set (II) (bottom) at $T=0$. At lower $\mu$ regime, $\mu\lesssim 0.28$ GeV, all the diquark masses are simply evaluated by linear functions as $m_{[qq]_\pm} = m_{[qq]_\pm}^{\rm vac}-2\mu$ and $m_{[sq]_\pm} = m_{[sq]_\pm}^{\rm vac}-2\mu$, since in this region medium effects do not enter and the diquark masses are diminished by the chemical potential solely. Beyond $\mu\sim 0.28$ GeV the diquark masses behave nonlinearly as $\mu$ increases accompanied by the medium effects, but soon they reach a critical chemical potential $\mu^*$ at which $m_{[qq]_+}=0$ is satisfied. Within our present NJL model, the $\mu^*$ represents the onset of the two-flavor color superconductivity (2SC) phase as explicitly shown in Appendix~\ref{sec:2SC}~\cite{Buballa:2003qv,Alford:2007xm}.\footnote{In two-color QCD where diquarks become color-singlet baryons, similarly to our present work, it is shown that the critical chemical potential $\mu^*$ which denotes the onset of emergence of the diquark condensates (i.e., the onset of the baryon superfluidity phase) is determined at which the $0^+$ diquark mass vanishes~\cite{Kogut:1999iv,Kogut:2000ek,Ratti:2004ra,Suenaga:2022uqn}. Moreover, it is supported by the lattice QCD simulations numerically~\cite{Hands:2001ee,Braguta:2016cpw,Iida:2019rah}.} In the 2SC phase, for instance, a positive-parity diquark $(\eta_+)_{i=3}^{a=3}$ defined in Eq.~(\ref{Eta+-}) creates a Bose-Einstein condensate (BEC), and accordingly, the original baryon (quark) number symmetry is violated such that the rotated baryon (quark) number symmetry becomes realized alternatively. As a result, mixings among diquarks and mesons sharing the identical quantum numbers, e.g., $f_0$ mesons (scalar and isoscalar mesons), $[qq]_+$ diquark and $[\bar{q}\bar{q}]_+$ antidiquark, occur~\cite{Blaschke:2004cs,Ebert:2004dr}. Thus, although the realization or indication of chiral-partner structures at $\mu\gtrsim \mu^*$ seems to be shown in Fig.~\ref{fig:DMass_T0}, they may include ambiguities. We leave investigation of the chiral-partner structures in cold and dense matter including the 2SC phase to future studies.

\section{Discussions}
\label{sec:Discussions}

\subsection{Decay of $\Lambda_c(1/2^-)$ at finite temperature}
\label{sec:Decays}

So far, we have focused on the masses of diquarks in medium which are not direct observable due to the color confinement. Useful testing grounds to see the diquark dynamics are color-singlet hadrons composed of a diquark and heavy quarks, such as SHBs and doubly heavy tetraquarks. In this section, we examine the masses and decay widths of the SHBs based on the analysis done for the diquarks in Sec.~\ref{sec:Results}. 

The singly charmed baryons composed of $[qq]_\pm$ are the ground-state $\Lambda_c(2286)$ and its chiral partner $\Lambda_c(1/2^-)$. Experimentally, the chiral partner $\Lambda_c(1/2^-)$ has not been identified, while the ground-state $\Lambda_c(2286)$ is well established~\cite{Workman:2022ynf}. One possible reason why the $\Lambda_c(1/2^-)$ is still missing could be a too large decay width caused by a comparably large mass of $\Lambda_c(1/2^-)$. Based on this speculation here we particularly focus on the mass and decay width of $\Lambda_c(1/2^-)$ at finite temperature.

We here assume the masses of $\Lambda_c(1/2^\pm)$ are given simply by the sum of the constituent $c$ quark mass $m_Q$ and the corresponding diquark ones $m_{[qq]_\pm}$ as~\cite{Harada:2019udr} 
\begin{eqnarray}
 M[\Lambda_c(1/2^\pm)] = m_Q + m_{[qq]_\pm}\ ,
\end{eqnarray}
based on the heavy-quark effective theory. From the particle data group (PDG) we find $M[\Lambda_c(1/2^+)] = 2.286$ GeV~\cite{Workman:2022ynf}. Hence, when we assume $m_{[qq]_+} = 0.725$ GeV in the vacuum as shown in Table~\ref{tab:DiquarkMass}, the value of $m_Q$ is fixed to be $m_Q=1.56$ GeV. With this value, $ M[\Lambda_c(1/2^-)]$ in the vacuum is evaluated to be $ M[\Lambda_c(1/2^-)]=2.99$ GeV when we take  the Set (II) in Table~\ref{tab:DiquarkMass}. The estimated mass is sufficiently larger than $ M[\Lambda_c(1/2^+)]$, which seems to be suitable for the demonstration in this section. For this reason, in the following analysis we will employ the Set (II) to study the mass and decay width of $\Lambda_c(1/2^-)$.\footnote{As for the mass of $\Lambda_c(1/2^-)$, for instance, the nonrelativistic quark model predicts $M[\Lambda_c(1/2^-)]=2.89$ GeV~\cite{Yoshida:2015tia}. Within the present chiral-model approach such a mass value is obtained when the inverse mass hierarchy is realized as with the Set (II).}

\begin{figure}[t]
\centering
\hspace*{-0.5cm} 
\includegraphics*[scale=0.8]{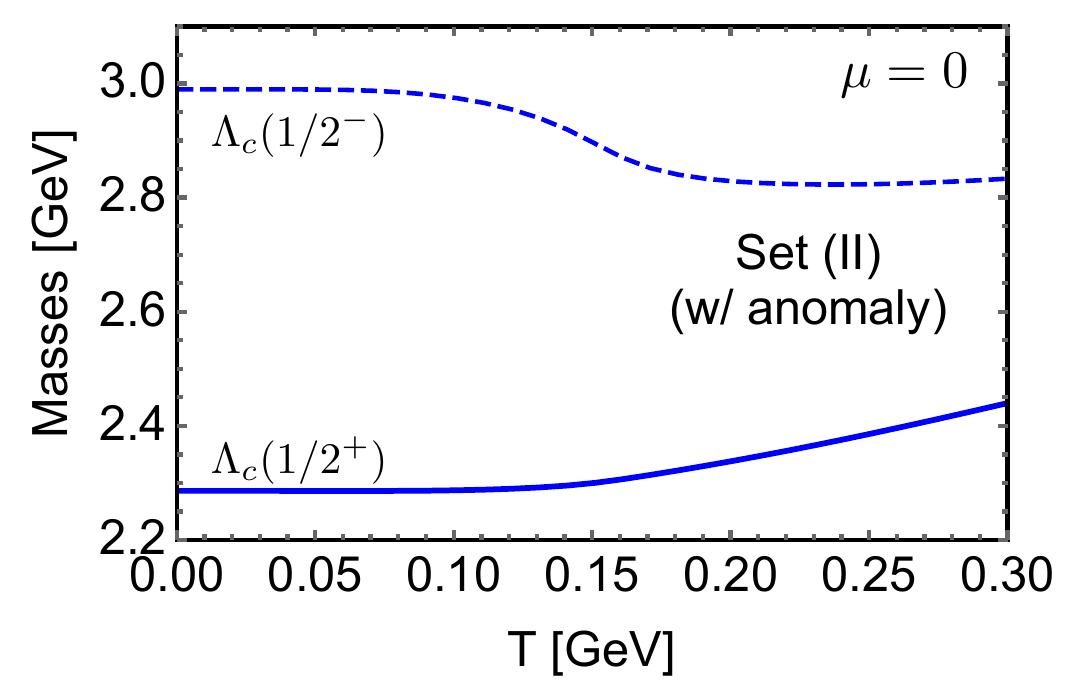}
\caption{Temperature dependence of the masses of $\Lambda_c(1/2^+)$ and $\Lambda_c(1/2^-)$ for $\mu=0$ with the Set (II).}
\label{fig:LcMass}
\end{figure}

\begin{figure}[t]
\centering
\hspace*{-0.5cm} 
\includegraphics*[scale=0.8]{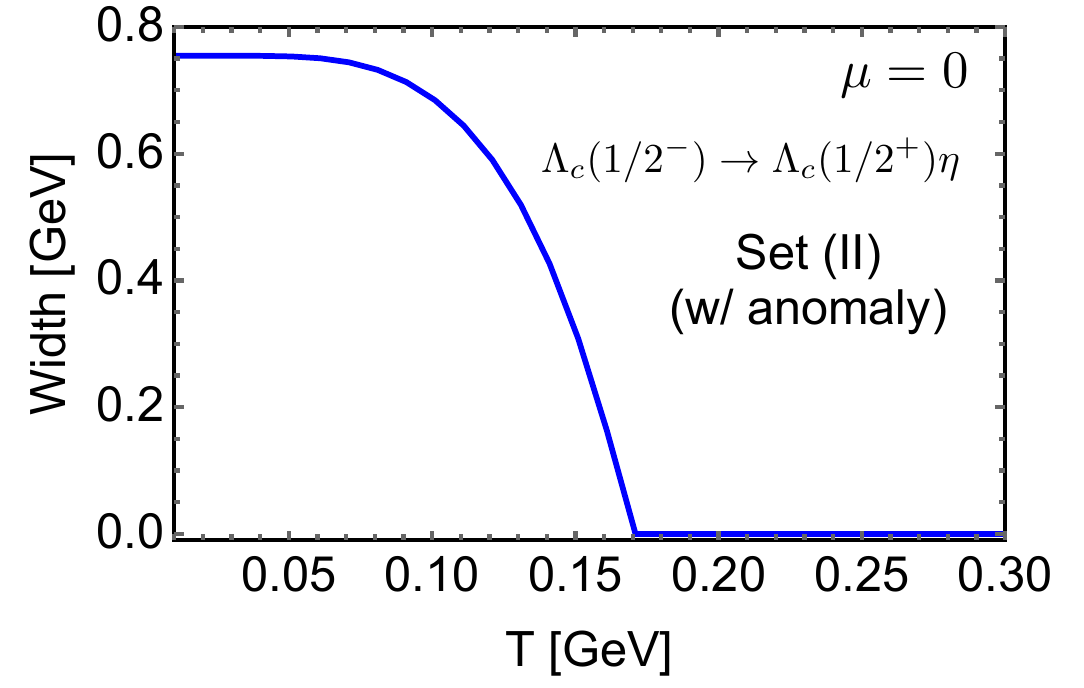}
\caption{Temperature dependence of the decay width of $\Lambda_c(1/2^-)\to\Lambda_c(1/2^+)\eta$ for $\mu=0$ with the Set (II).}
\label{fig:LcDecay}
\end{figure}

Under an assumption that the value of $m_Q$ and interactions between the heavy quark and the diquark do not change in medium, the temperature dependence of the masses of $\Lambda_c(1/2^\pm)$ is simply evaluated by that of $m_{[qq]_\pm}$. Depicted in Fig.~\ref{fig:LcMass} is the resultant temperature dependences of $M[\Lambda_c(1/2^\pm)]$ for $\mu=0$ with the Set (II). The dependences are essentially identical to those of $[qq]_\pm$ as in the top-right panel of Fig.~\ref{fig:DMass_Mu0}, implying that the mass of $\Lambda_c(1/2^+)$ increases while that of $\Lambda_c(1/2^-)$ decreases as the system is heated.

The main decay mode of $\Lambda_c(1/2^-)$ is $\Lambda_c(1/2^-)\to\Lambda_c(1/2^+)\eta$ due to $SU(2)_I$ isospin symmetry. In the previous analysis based on the linear representation of $SU(3)_L\times SU(3)_R$ chiral symmetry, which is basically equivalent to our present approach~\cite{Kawakami:2020sxd}, it was found that couplings of $\Lambda_c(1/2^-)$-$\Lambda_c(1/2^+)$-$\eta$ are given by a generalized Goldberger-Treiman relation as
\begin{eqnarray}
{\cal L}_{\rm int} &=&  \frac{2i}{\sqrt{3}f_\pi}\Bigg[M_{B1}\left(\cos\theta_P + \frac{1}{\sqrt{2}}\sin\theta_P\right) + \nonumber\\
&& M_{B2}\left(\cos\theta_P-\sqrt{2}\sin\theta_P\right)\Bigg]\eta\bar{\Lambda}_c(1/2^+)\Lambda_c(1/2^-) \ . \nonumber\\ \label{Coupling}
\end{eqnarray}
Here, $M_{B1}$ and $M_{B2}$ are evaluated by mass differences between the chiral partners as
\begin{eqnarray}
M_{B1} &=& \frac{A\Delta M_{\Xi_c}-\Delta M_{\Lambda_c}}{2(A^2-1)}\ , \nonumber\\
M_{B2} &=& \frac{A\Delta M_{\Lambda_c}-\Delta  M_{\Xi_c}}{2(A^2-1)}\ , \label{MB1B2}
\end{eqnarray}
with
\begin{eqnarray}
\Delta M_{\Lambda_c} &\equiv& M[\Lambda_c(1/2^-)]-M[\Lambda_c(1/2^+)]\ , \nonumber\\
\Delta M_{\Xi_c} &\equiv& M[\Xi_c(1/2^-)]-M[\Xi_c(1/2^+)]\  , \label{DeltaSHB}
\end{eqnarray}
and the mixing angle $\theta_P=-11.3^\circ$ appears due to the $\eta$-$\eta'$ mixing. The dimensionless constant $A$ in Eq.~(\ref{MB1B2}) quantifies a violation of $SU(3)_{L+R}$ flavor symmetry, which is estimated in our present approach as $A = (2f_K-f_\pi)/f_\pi=1.39$. We note that $M[\Xi_c(1/2^\pm)]$ in Eq.~(\ref{DeltaSHB}) is the masses of $\Xi_c(1/2^\pm)$ composed of a $c$ quark and a $[sq]_\pm$ diquark, which is defined similarly to $M[\Lambda_c(1/2^\pm)]$.

From the coupling~(\ref{Coupling}) the decay width of $\Lambda_c(1/2^-)\to\Lambda_c(1/2^+)\eta$ is computed, and the resultant temperature dependence of the width is displayed in Fig.~\ref{fig:LcDecay}. In obtaining the figure, the value of $\eta$ mass is fixed to be $m_\eta=0.548$ GeV~\cite{Workman:2022ynf}. Besides, any thermal effects such as the broadening effect are neglected, and only the changes of $M[\Lambda_c(1/2^\pm)]$ and $M[\Xi_c(1/2^\pm)]$ at finite temperature are incorporated through Eq.~(\ref{MB1B2}). The figure indicates that the decay width vanishes above $T_{\rm pc}$ since the threshold is closed ($\Delta M_{\Lambda_c}<m_\eta$) although the width is exceedingly large in the vacuum ($\Delta M_{\Lambda_c}> m_\eta$).

Our demonstration in this section implies that, even though it is difficult to observe $\Lambda_c(1/2^-)$ in the vacuum due to its too large decay width, there would be a possibility of observing $\Lambda_c(1/2^-)$ when focusing on finite-temperature system by e.g., HICs or lattice simulations. Toward a realistic evaluation, thermal effects such as the broadening effects are unavoidable, and we leave inclusion of such effects for a future study. 

Moreover, other decay modes of $\Lambda_c(1/2^-)$ are expected. As a primary mode among them, $\Lambda_c(1/2^-)$ would decay into the ground-state $\Lambda_c(1/2^+)$ by emitting two pions sequentially via $\Sigma_c$ resonances~\cite{Kawakami:2020sxd}. However, such processes break heavy-quark spin symmetry so that those channels are rather suppressed. In addition, in Ref.~\cite{Kim:2022pyq} it was predicted that masses of $\Sigma_c$'s decrease in accordance with the partial restoration of chiral symmetry, which leads to a closing of $\Sigma_c\to \Lambda_c(1/2^+)\pi$ channel. Therefore, we expect that the sequential decays of $\Lambda_c(1/2^-)\to\Sigma_c\pi \to \Lambda_c(1/2^+)\pi\pi$ do not generate sizable widths even at finite temperature.

\subsection{Artifacts from of the proper-time regularization}
\label{sec:Artifact}

As shown in Sec.~\ref{sec:Regularization}, in our present analysis the three-dimensional proper-time regularization including UV and IR cutoffs is employed so as to evaluate the mass of diquarks in a transparent way by removing the imaginary parts. In this section, we discuss the appearance of artifacts which would break chiral symmetry in the chiral limit.

From Eq.~(\ref{FPi}), the pion decay constant in the chiral limit $\bar{f}_\pi$ is evaluated to be
\begin{eqnarray}
\bar{f}_\pi &=& 3\sqrt{2\bar{Z}_\pi}\bar{M}_q\int\frac{d^3p}{(2\pi)^3}\frac{1}{\bar{E}_{\bm p}} \nonumber\\
&\times& \left(\frac{{\rm e}^{-\frac{2\bar{E}_{\bm p}}{\Lambda_{\rm UV}}}-{\rm e}^{-\frac{2\bar{E}_{\bm p}}{\mu_{\rm IR}}}}{2\bar{E}_{\bm p}^2} + \frac{\frac{1}{\Lambda_{\rm UV}}{\rm e}^{-\frac{2\bar{E}_{\bm p}}{\Lambda_{\rm UV}}}-\frac{1}{\mu_{\rm IR}}{\rm e}^{-\frac{2\bar{E}_{\bm p}}{\mu_{\rm IR}}}}{\bar{E}_{\bm p}}\right) \nonumber\\
&\times& \Big[1-f_F\left(\bar{\epsilon}_{\rm p}({\bm p})\right)-f_F\left(\bar{\epsilon}_{\rm a}({\bm p})\right)\Big]\ , \label{FPiLim}
\end{eqnarray}
by taking a limit of $m_\pi\to0$. In Eq.~(\ref{FPiLim}), we have defined the dynamical quark mass $\bar{M}_q$ and dispersion relations $\bar{\epsilon}_\zeta({\bm p})$ in the chiral limit as
\begin{eqnarray}
\bar{M}_q = -4G\langle\bar{q}q\rangle + 2K\langle\bar{q}q\rangle^2\ , \label{MassLim}
\end{eqnarray}
and
\begin{eqnarray}
\bar{\epsilon}_\zeta({\bm p}) &=& \bar{E}_{\bm p}-\eta_\zeta\mu\ ,
\end{eqnarray}
respectively, with $\bar{E}_{\bm p} = \sqrt{{\bm p}^2+\bar{M}_q^2}$. It should be noted that $\langle\bar{q}q\rangle$ in Eq.~(\ref{MassLim}) is the chiral condensate evaluated in the chiral limit. The quantity $\bar{Z}_\pi$ in Eq.~(\ref{FPiLim}) is the renormalization constant for the pion in the chiral limit, which is expressed as
\begin{eqnarray}
\bar{Z}_\pi^{-1}&=& 3\int\frac{d^3p}{(2\pi)^3}\frac{1}{\bar{E}_{\bm p}}\Bigg(\frac{{\rm e}^{-\frac{2\bar{E}_{\bm p}}{\Lambda_{\rm UV}}}-{\rm e}^{-\frac{2\bar{E}_{\bm p}}{\mu_{\rm IR}}}}{2\bar{E}_{\bm p}^2} \nonumber\\
&+& \frac{\frac{1}{\Lambda_{\rm UV}}{\rm e}^{-\frac{2\bar{E}_{\bm p}}{\Lambda_{\rm UV}}}-\frac{1}{\mu_{\rm IR}}{\rm e}^{-\frac{2\bar{E}_{\bm p}}{\mu_{\rm IR}}}}{\bar{E}_{\bm p}} \nonumber\\
&+& \frac{1}{\Lambda_{\rm UV}^2}{\rm e}^{-\frac{2\bar{E}_{\bm p}}{\Lambda_{\rm UV}}}-\frac{1}{\mu_{\rm IR}^2}{\rm e}^{-\frac{2\bar{E}_{\bm p}}{\mu_{\rm IR}}}\Bigg) \nonumber\\
&\times& \Big[1-f_F\left(\bar{\epsilon}_{\rm p}({\bm p})\right)-f_F\left(\bar{\epsilon}_{\rm a}({\bm p})\right)\Big]\ ,
\end{eqnarray}
from Eq.~(\ref{ZPi}). That is, within our proper-time regularization scheme the decay constant and the dynamical quark mass are related as
\begin{eqnarray}
\bar{f}_\pi = \sqrt{2}\bar{Z}_\pi^{-1/2}\bar{M}_q+\delta \bar{f}_\pi \ ,\label{FPiZPiLim}
\end{eqnarray}
with
\begin{eqnarray}
\delta \bar{f}_\pi &\equiv& -3\sqrt{2\bar{Z}_\pi}\bar{M}_q\int\frac{d^3p}{(2\pi)^3} \frac{1}{\Lambda_{\rm UV}^2}{\rm e}^{-\frac{2\bar{E}_{\bm p}}{\Lambda_{\rm UV}}}-\frac{1}{\mu_{\rm IR}^2}{\rm e}^{-\frac{2\bar{E}_{\bm p}}{\mu_{\rm IR}}} \nonumber\\
&\times&\Big[1-f_F\left(\epsilon_{\rm p}^{(n)}({\bm p})\right)-f_F\left(\epsilon_{\rm a}^{(n)}({\bm p})\right)\Big] \ . \label{DeltaFPi}
\end{eqnarray}

On the other hand, we know that those quantities must satisfy $\bar{f}_\pi=\sqrt{2}\bar{Z}_\pi^{-1/2}\bar{M}_q$ from the Glashow-Weinberg relation due to exact chiral symmetry~\cite{Glashow:1967rx}, and thus the existence of Eq.~(\ref{DeltaFPi}) implies an artificial violation of chiral symmetry. Such a troublesome contributions stem from pion mass dependences in the exponents of $\bar{Z}_\pi$ in Eq.~(\ref{ZPi}), which is obviously induced by the use of the proper-time regularization. However, the artifact~(\ref{DeltaFPi}) is proportional to $\bar{M}_q$ as the first term in Eq.~(\ref{FPiZPiLim}), so that the diquark masses at sufficiently high temperature where the chiral-symmetry restoration takes place well are not affected by the artificial violation significantly. Thus, our qualitative conclusion in this paper does not change. Besides, while $\langle\bar{s}s\rangle$ is not prominently reduced above $T_{\rm pc}$ as in Fig.~\ref{fig:Condensates}, the violation of chiral symmetry is dominantly triggered by the presence of the current $s$ quark mass $m_s$, and again it is expected that our main results are not affected by the artifact~(\ref{DeltaFPi}). Although our regularization breaks the Glashow-Weinberg relation, we note that the massless nature of a pion in the chiral limit can be checked as it should be.

We emphasize that the artifacts are not obtained as a direct consequence of the inclusion of IR cutoff $\mu_{\rm IR}$. In fact, the artifacts remain finite when we take $\mu_{\rm IR}\to0$ keeping $\Lambda_{\rm UV}$ finite in Eq.~(\ref{DeltaFPi}). Moreover, even the widely-used four-dimensional proper-time regularization with obvious Lorentz covariance in the vacuum suffers from similar artifacts.

\section{Conclusions}
\label{sec:Conclusions}

In this paper, we have investigated diquark masses at finite temperature and chemical potential based on the three-flavor NJL model from the viewpoint of the (partial) restoration of chiral symmetry and the $U(1)_A$ axial anomaly. In particular, we have focused on the mass degeneracies of the positive-parity and negative-parity diquarks at high temperature to see the chiral-partner structure. As a result, we have found that the inverse mass hierarchy caused by the $U(1)_A$ axial anomaly for the negative-parity diquarks remains valid at finite temperature. We have also found that the mass degeneracies take place clearly in all $[ud]$, $[su]$ and $[sd]$ diquark sectors in the absence of anomaly effects to the diquarks. On the other hand, the anomaly effect defers the mass degeneracy in $[ud]$ sector, reflecting the slow reduction of $\langle\bar{s}s\rangle$ at finite temperature, whereas those in $[su]$ and $[sd]$ sectors are manifestly realized reflecting the fast reduction of $\langle\bar{u}u\rangle$ and $\langle\bar{d}d\rangle$. Those findings are expected to provide future lattice simulations with useful information on the chiral-partner structure for the diquarks together with the magnitude of the $U(1)_A$ axial anomaly. 

As for low temperature and high density regime, our analysis indicates that the emergence of the color superconducting phase, especially the two-flavor superconductivity, is unavoidable toward delineation of the chiral-partner structures of diquarks. Thus, we leave examination of diquarks in the color superconducting phase for a future study. 

Besides, based on the temperature dependence of diquark masses, we have discussed decay widths of $\Lambda_c(1/2^-)$, the chiral partner of $\Lambda_c(2286)$ which has not been experimentally observed, at finite temperature. As a result, we have found that the decay channel of $\Lambda_c(1/2^-)$ is closed accompanied by the partial restoration of chiral symmetry, which would demonstrate a possibility of observing the missing $\Lambda_c(1/2^-)$ in future HIC experiments. To check such a feasibility, more realistic evaluations including the broadening effects are inevitable and we leave such study for future publication.

\section*{acknowledgment}
The authors thank Masayasu Harada and Daisuke Jido for useful comments and discussions. This work was supported by the RIKEN special postdoctoral researcher program (D.S.), and by Grants-in-Aid for Scientific Research No.~JP20K03959 and No.~JP21H00132 (M.O.) from Japan Society for the Promotion of Science.

\appendix

\section{Evaluation of $m_\pi$, $m_K$, $f_\pi$ and $f_K$ }
\label{sec:PSInput}

In this appendix, we give explanations how to evaluate the pion mass $m_\pi$, kaon mass $m_K$, pion decay constant $f_\pi$ and kaon decay constant $f_K$ in the vacuum.

Fist, we calculate the pion mass $m_\pi$ and kaon mass $m_K$. Similarly to the diquark masses given in the main text, those masses are also evaluated by pole positions of the corresponding BS amplitudes
\begin{eqnarray}
{\cal T} = (1-{\cal K}{\cal J})^{-1}{\cal K} \ . \label{BSPiK}
\end{eqnarray}
Kernels for pion and kaon channels are read off from effective four-point interactions of quarks in Eq.~(\ref{LNJL}) with the approximation~(\ref{Approximation}), yielding
\begin{eqnarray}
{\cal K}_{\pi} &=& i(4G-2K\langle\bar{s}s\rangle)\ , \nonumber\\
{\cal K}_K &=& i(4G-2K\langle\bar{q}q\rangle) \ , \label{KPiK}
\end{eqnarray}
respectively. Besides, the respective loop functions read
\begin{eqnarray}
{\cal J}_\pi &=& -3\int\frac{d^4p}{(2\pi)^4}{\rm tr}\Big[i\gamma_5S_{(q)}(p')i\gamma_5 S_{(q)}(p)\Big] \ , \nonumber\\
{\cal J}_{K} &=& -3\int\frac{d^4p}{(2\pi)^4}{\rm tr}\Big[i\gamma_5S_{(q)}(p')i\gamma_5 S_{(s)}(p)\Big] \ .
\end{eqnarray}
Then, using the Dirac trace formula~(\ref{DiracTrace}) and performing the $q_0$ integral, we arrive at
\begin{eqnarray}
&& {\cal J}_\pi(q_0) \nonumber\\
&& = 6i\int\frac{d^3p}{(2\pi)^3}\Bigg\{ \frac{{\cal R}\big(q_0-2E^{(q)}_{\bm p}\big)}{q_0-2E^{(q)}_{\bm p}} - \frac{{\cal R}\big(q_0+2E^{(q)}_{\bm p}\big)}{q_0+2E^{(q)}_{\bm p}} \Bigg\}\ , \label{JPi} \nonumber\\
\end{eqnarray}
and
\begin{eqnarray}
&&{\cal J}_{{K}} (q_0) \nonumber\\
&&= 3i\int\frac{d^3p}{(2\pi)^3} \left(1+\frac{{\bm p}^2+M_sM_q}{E_{\bm p}^{(s)}E_{\bm p}^{(q)}}\right) \nonumber\\
&&\times\Bigg\{\frac{{\cal R}\big(q_0-E_{\bm p}^{(s)}-E_{\bm p}^{(q)}\big)}{q_0-E_{\bm p}^{(s)}-E_{\bm p}^{(q)}} - \frac{{\cal R}\big(q_0+E_{\bm p}^{(s)}+E_{\bm p}^{(q)}\big)}{q_0+E_{\bm p}^{(s)}+E_{\bm p}^{(q)}} \ \Bigg\} \ , \label{JK}\nonumber\\
\end{eqnarray}
at rest frame ${\bm q}={\bm 0}$. In Eqs.~(\ref{JPi}) and~(\ref{JK}) we have employed the same regularization technique as the diquark loop functions so as to maintain the chiral symmetric consistency. Inserting the kernels~(\ref{KPiK}) and the loop functions~(\ref{JPi}) and~(\ref{JK}) into the BS amplitude~(\ref{BSPiK}), $m_\pi$ and $m_K$ are computed.

Next, we present analytic expressions of the decay constants $f_\pi$ and $f_K$. The decay constants are defined through matrix elements of
\begin{eqnarray}
&& \langle 0|\bar{\psi}\gamma^\mu\gamma_5(\lambda_f^a/2)\psi|\pi^b(q)\rangle = -if_\pi q^\mu\ \ \ \ (a=b=1-3)\ , \nonumber\\
&&  \langle 0|\bar{\psi}\gamma^\mu\gamma_5(\lambda_f^a/2)\psi|K^b(q)\rangle = -if_K q^\mu\ \ \ \ (a=b=4-7)\ , \nonumber\\
\end{eqnarray} 
and in our present normalization they are computed as
\begin{widetext}
\begin{eqnarray}
f_\pi &=&  -\frac{3i\sqrt{Z_\pi}}{\sqrt{2} q^2}\int\frac{d^4p}{(2\pi)^4}{\rm tr}[\Slash{q}\gamma_5S_{(q)}(p+q)\gamma_5S_{(q)}(p)]\Big|_{q_0=m_\pi,{\bm q}={\bm 0}}  \nonumber\\
&=& -\frac{3\sqrt{2Z_\pi}M_n}{ m_\pi} \int\frac{d^3p}{(2\pi)^3}\frac{1}{E_{\bm p}^{(q)}}\left(\frac{{\cal R}\big(m_\pi-2E_{\bm p}^{(q)}\big)}{m_\pi-2E_{\bm p}^{(q)}}   + \frac{{\cal R}\big(m_\pi+2E_{\bm p}^{(q)}\big)}{m_\pi+2E_{\bm p}^{(q)}}  \right) \ , \label{FPi}
\end{eqnarray}
and
\begin{eqnarray}
f_K &=& -\frac{3i\sqrt{Z_{K}}}{\sqrt{2} q^2}\int\frac{d^4p}{(2\pi)^4}{\rm tr}\left[\Slash{q}\gamma_5S_{(q)}(p+q)\gamma_5S_{(s)}(p)\right]\Big|_{q_0=m_{{K}},{\bm q}={\bm 0}} \nonumber\\
&=& -\frac{3\sqrt{Z_{{K}}}}{\sqrt{2} m_{{K}}}\int\frac{d^3p}{(2\pi)^3}\left(\frac{M_q}{E_{\bm p}^{(q)}}+\frac{M_s}{E_p^{(s)}}\right) \Bigg\{\frac{{\cal R}\big(m_{{K}}-E_{\bm p}^{(q)}-E_{\bm p}^{(s)}\big)}{m_{{K}}-E_{\bm p}^{(q)}-E_{\bm p}^{(s)}} +\frac{{\cal R}\big(m_{{K}}+E_{\bm p}^{(q)}+E_{\bm p}^{(s)}\big)}{m_{{K}}+E_{\bm p}^{(q)}+E_{\bm p}^{(s)}}  \Bigg\}\ . \label{FK}
\end{eqnarray}
In these expressions $Z_\pi$ and $Z_K$ are renormalization constants for pion and kaon wave functions, respectively, which are defined by 
\begin{eqnarray}
Z_\pi^{-1} &\equiv& \frac{i}{2m_\pi}\frac{\partial {\cal J}_\pi(q_0)}{\partial q_0}\Big|_{q_0=m_\pi} =  -\frac{3}{m_\pi}\int\frac{d^3p}{(2\pi)^3}\Big[{\cal F}\big(m_\pi-2E_{\bm p}^{(q)}\big) -{\cal F}\big(m_\pi+2E_{\bm p}^{(q)}\big)\Big]\ , \label{ZPi}
\end{eqnarray}
and
\begin{eqnarray}
Z_K^{-1} &\equiv& \frac{i}{2m_K}\frac{\partial {\cal J}_K(q_0)}{\partial q_0}\Big|_{q_0=m_K} \nonumber\\
&=& -\frac{3}{2m_{{K}}}\int\frac{d^3p}{(2\pi)^3} \left(1+\frac{{\bm p}^2+M_sM_q}{E_{\bm p}^{(s)}E_{\bm p}^{(q)}}\right)  \Bigg\{{\cal F}\big(m_K-E_{\bm p}^{(q)}-E_{\bm p}^{(q)}\big) - {\cal F}\big(m_K+E_{\bm p}^{(q)}+E_{\bm p}^{(s)}\big) \Bigg\} \ ,
\end{eqnarray}
with
\begin{eqnarray}
{\cal F}(x) \equiv \frac{\partial}{\partial x}\left(\frac{{\rm e}^{-\frac{|x|}{\Lambda_{\rm UV}}}-{\rm e}^{-\frac{|x|}{\mu_{\rm IR}}}}{x} \right) = \frac{{\rm e}^{-\frac{|x|}{\mu_{\rm IR}}}-{\rm e}^{-\frac{|x|}{\Lambda_{\rm UV}}}}{x^2} + \frac{\frac{1}{\mu_{\rm IR}}{\rm e}^{-\frac{|x|}{\mu_{\rm IR}}}-\frac{1}{\Lambda_{\rm UV}}{\rm e}^{-\frac{|x|}{\Lambda_{\rm UV}}}}{|x|}\ . \label{CapFX}
\end{eqnarray}  
\end{widetext}
At first glance, the function ${\cal F}(x)$ seems to yield a discontinuity at $x=0$ originating from derivatives of $|x|$ with respect to $x$. However, one can easily show
\begin{eqnarray}
\lim_{x\to+0}{\cal F}(x) = \lim_{x\to-0}{\cal F}(x) = \frac{1}{2\Lambda_{\rm UV}^2}-\frac{1}{2\mu_{\rm IR}^2}\ ,
\end{eqnarray}
and no such discontinuities emerge. Therefore, the renormalization constants are well defined in our treatment.

\section{Emergence of the 2SC phase}
\label{sec:2SC}

Here, we analytically show that the onset density of 2SC phase is estimated when $m_{[qq]_+}$ becomes zero within our present model.

The 2SC phase is defined by emergence of diquark condensates made of $u$ and $d$ quarks. In particular, the condensates are $S$-wave, flavor-singlet and color anti-triplet~\cite{Buballa:2003qv,Alford:2007xm}, so that the diquark gap takes the form of, e.g.,
\begin{eqnarray}
\Delta_{\rm 2SC}  \equiv \sqrt{2} \langle(\eta_+)_{i=3}^{a=3}\rangle = -\frac{1}{2}\langle q^{T} C\gamma_5\tau_f^2\lambda_c^2 q\rangle \ , \label{2SCDelta}
\end{eqnarray}
where $\tau_f^A$ is the Pauli matrix acting on two-flavor $q=(u,d)^T$ space. Including the diquark condensate~(\ref{2SCDelta}) in addition to the chiral condensates $\langle\bar{q}q\rangle$ and $\langle\bar{s}s\rangle$, at mean-field level the Lagrangian~(\ref{LNJL}) is reduced to
\begin{eqnarray}
{\cal L}_{\rm MF} &=&  \bar{q}(i\Slash{\partial}+\mu\gamma_0-M_q)q + \bar{s}(i\Slash{\partial}+\mu\gamma_0 -M_s)s \nonumber\\
&-& \left(H-\frac{K'}{4}\langle\bar{s}s\rangle\right)\left(\Delta^*_{\rm 2SC}q^{T}\tau^2_f\lambda^2_cC\gamma_5q+{\rm h.c.}\right) \nonumber\\
&-& 2G(2\langle\bar{q}q\rangle^2+\langle\bar{s}s\rangle^2)-4H|\Delta_{\rm 2SC}|^2 \nonumber\\
&+& 4K\langle\bar{q}q\rangle^2\langle\bar{s}s\rangle + 2K'|\Delta_{\rm 2SC}|^2\langle\bar{s}s\rangle \ , \label{LMF2SC}
\end{eqnarray}
where dynamical quark masses can be now affected by the diquark condensate $\Delta_{\rm 2SC}$ as
\begin{eqnarray}
M_q &=& m_q- 4G\langle\bar{q}q\rangle+2K\langle\bar{q}q\rangle\langle\bar{s}s\rangle\ , \nonumber\\
M_s &=& m_s- 4G\langle\bar{s}s\rangle + 2K\langle\bar{q}q\rangle^2 + K'|\Delta_{\rm 2SC}|^2\ .
\end{eqnarray}
From the mean-field Lagrangian~(\ref{LMF2SC}), a thermodynamic potential per volume $V$ is evaluated to be ($\beta=1/T$)
\begin{widetext}
\begin{eqnarray}
\Omega/V &=& -8\int\frac{d^3p}{(2\pi)^3}\left[\frac{\tilde{\epsilon}^{(q)}_{\rm p}({\bm p})}{2} + \frac{\tilde{\epsilon}^{(q)}_{\rm a}({\bm p})}{2} + T{\rm ln}\left(1+{\rm e}^{-\beta\tilde{\epsilon}^{(q)}_{\rm p}({\bm p})}\right)+ T{\rm ln}\left(1+{\rm e}^{-\beta\tilde{\epsilon}^{(q)}_{\rm a}({\bm p})}\right)\right] \nonumber\\
&-& 4\int\frac{d^3p}{(2\pi)^3}\left[\frac{{\epsilon}^{(q)}_{\rm p}({\bm p})}{2} + \frac{{\epsilon}^{(q)}_{\rm a}({\bm p})}{2} + T{\rm ln}\left(1+{\rm e}^{-\beta{\epsilon}^{(q)}_{\rm p}({\bm p})}\right)+ T{\rm ln}\left(1+{\rm e}^{-\beta{\epsilon}^{(q)}_{\rm a}({\bm p})}\right)\right] \nonumber\\
&-& 6\int\frac{d^3p}{(2\pi)^3}\left[\frac{{\epsilon}^{(s)}_{\rm p}({\bm p})}{2} + \frac{{\epsilon}^{(s)}_{\rm a}({\bm p})}{2} + T{\rm ln}\left(1+{\rm e}^{-\beta{\epsilon}^{(s)}_{\rm p}({\bm p})}\right)+ T{\rm ln}\left(1+{\rm e}^{-\beta{\epsilon}^{(s)}_{\rm a}({\bm p})}\right)\right] \nonumber\\
&+& 2G(2\langle\bar{q}q\rangle^2+\langle\bar{s}s\rangle^2) +4H|\Delta_{\rm 2SC}|^2 - 4K\langle\bar{q}q\rangle^2\langle\bar{s}s\rangle - 2K'|\Delta_{\rm 2SC}|^2\langle\bar{s}s\rangle\ , \label{Omega2SC}
\end{eqnarray}
\end{widetext}
where dispersion relations of $q$ and $s$ quarks are given by ($\zeta={\rm p},{\rm a}$)
\begin{eqnarray}
\epsilon_\zeta^{(f)} = E_\zeta^{(f)}-\eta_\zeta\mu\ , 
\end{eqnarray}
with $E_{\bm p}^{(f)} = \sqrt{{\bm p}^2+M_f^2}$, and those of quasiparticles corrected by the diquark condensate read
\begin{eqnarray}
\tilde{\epsilon}_\zeta^{(q)}({\bm p}) = \sqrt{\left({\epsilon}_\zeta^{(q)}({\bm p})\right)^2+(4H-K'\langle\bar{s}s\rangle)^2|\Delta_{\rm 2SC}|^2}\ . 
\end{eqnarray}
The factors for each quark contribution in Eq.~(\ref{Omega2SC}) are understood by (spin)$\times$(flavor)$\times$(color) degrees of freedom.

From a stationary condition of Eq.~(\ref{Omega2SC}) with respect to $\Delta_{\rm 2SC}$, a gap equation determining the value of diquark condensate $\Delta_{\rm 2SC}$ in the 2SC phase is obtained as
\begin{widetext}
\begin{eqnarray}
0 &=& 2(4H-K'\langle\bar{s}s\rangle)^2\int\frac{d^3p}{(2\pi)^3}\Bigg\{\frac{{\cal R}\big(2\tilde{\epsilon}^{(q)}_{\rm p}({\bm p})\big)}{2\tilde{\epsilon}^{(q)}_{\rm p}({\bm p})} \Big[1-2f_F\big(\tilde{\epsilon}^{(q)}_{\rm p}({\bm p})\big)\Big] + \frac{{\cal R}\big(2\tilde{\epsilon}^{(q)}_{\rm a}({\bm p})\big)}{2\tilde{\epsilon}^{(q)}_{\rm a}({\bm p})}\Big[1-2f_F\big(\tilde{\epsilon}^{(q)}_{\rm a}({\bm p})\big)\Big]\Bigg\} \nonumber\\
&+& 6K'M_s\int\frac{d^3p}{(2\pi)^3}\frac{{\cal R}\big(2E_{\bm p}^{(s)}\big)}{2E_{\bm p}^{(s)}}\left\{1-f_F\big({\epsilon}^{(s)}_{\rm p}({\bm p})\big)-f_F\big({\epsilon}^{(s)}_{\rm a}({\bm p})\big)\right\} -2H + K'\langle\bar{s}s\rangle\ , \label{Gap2SC}
\end{eqnarray}
and thus, by taking $\Delta_{\rm 2SC}\to0$ in Eq.~(\ref{Gap2SC}), an identity which holds at the onset density of 2SC phase is found to be
\begin{eqnarray}
0 &=&2 (4H-K'\langle\bar{s}s\rangle)^2\int\frac{d^3p}{(2\pi)^3}\Bigg\{\frac{{\cal R}\big(2{\epsilon}^{(q)}_{\rm p}({\bm p})\big)}{2{\epsilon}^{(q)}_{\rm p}({\bm p})}  \Big[1-2f_F\big({\epsilon}^{(q)}_{\rm p}({\bm p})\big)\Big] + \frac{{\cal R}\big(2{\epsilon}^{(q)}_{\rm a}({\bm p})\big)}{2{\epsilon}^{(q)}_{\rm a}({\bm p})}\Big[1-2f_F\big({\epsilon}^{(q)}_{\rm a}({\bm p})\big)\Big]\Bigg\} \nonumber\\
&+& 6K'M_s\int\frac{d^3p}{(2\pi)^3}\frac{{\cal R}\big(2E_{\bm p}^{(s)}\big)}{2E_{\bm p}^{(s)}}\left\{1-f_F\big({\epsilon}^{(s)}_{\rm p}({\bm p})\big)-f_F\big({\epsilon}^{(s)}_{\rm a}({\bm p})\big)\right\} - 2H + K'\langle\bar{s}s\rangle\ . \label{Gap2SC0}
\end{eqnarray}
This identity is further reduced; from the analytic expression of $\langle\bar{s}s\rangle$ in Eq.~(\ref{SSAnalytic}), one can find that Eq.~(\ref{Gap2SC0}) yields
\begin{eqnarray}
1-2(4H-K'\langle\bar{s}s\rangle)\int\frac{d^3p}{(2\pi)^3}\left\{\frac{1}{{\epsilon}_{\rm p}^{(q)}({\bm p})}\Big[1-2f_F\big({\epsilon}^{(q)}_{\rm p}({\bm p})\big)\Big] + \frac{1}{{\epsilon}_{\rm a}^{(n)}({\bm p})}\Big[1-2f_F\big({\epsilon}^{(q)}_{\rm a}({\bm p})\big)\Big]\right\}= 0 \ , \label{Delta0I}
\end{eqnarray}
unless $4H=K'\langle\bar{s}s\rangle$. Meanwhile, from the kernel~(\ref{KernelD}) and the quark loop function~(\ref{JUD+}) for $[qq]_+$ diquark channel, we can see that the pole position of the BS amplitude~(\ref{TBSPole}) for this channel is determined by solving the following equation with respect to $q_0$:
\begin{eqnarray}
&& \delta^{ab}-{\cal K}^{ac}_{[qq]_+}{\cal J}^{cb}_{[qq]_+}(q_0) \nonumber\\
&=& \delta^{ab}+4\delta^{ab}(4H-K'\langle\bar{s}s\rangle)\int\frac{d^3p}{(2\pi)^3}\left\{\frac{1}{q_0-2{\epsilon}_{\rm p}^{(q)}({\bm p})}\Big[1-2f_F\big({\epsilon}^{(q)}_{\rm p}({\bm p})\big)\Big] - \frac{1}{q_0+2{\epsilon}_{\rm a}^{(n)}({\bm p})}\Big[1-2f_F\big({\epsilon}^{(q)}_{\rm a}({\bm p})\big)\Big]\right\}= 0\ . \nonumber\\ \label{PoleQQ+}
\end{eqnarray}
\end{widetext}
Therefore, from the identity~(\ref{Delta0I}), we can conclude that Eq.~(\ref{PoleQQ+}) has a solution when $q_0=0$, and it is shown that the onset of the 2SC phase is certainly triggered when $[qq]_+$ diquark mass becomes zero.

\bibliography{reference}

\end{document}